\documentclass[prfluids,groupedaddress,floatfix]{revtex4-2}

\usepackage{graphicx}
\usepackage[final]{changes}

\begin{document}


\title{The transient behaviour of through-flowing gravity currents interacting with a roughness array}


\author{Alex Meredith}
\author{Craig McConnochie}
\email{craig.mcconnochie@canterbury.ac.nz}
\author{Roger Nokes}
\affiliation{Department of Civil and Natural Resources Engineering, University of Canterbury, Christchurch, New Zealand}
\author{Claudia Cenedese}
\affiliation{Physical Oceanography Department, Woods Hole Oceanographic Institution, Woods Hole, MA, USA}


\date{\today}

\begin{abstract}
We present laboratory experiments that investigate the structure and flow characteristics of gravity currents travelling through an array of roughness elements.
The roughness elements are of comparable height to the gravity current such that the current flows through the roughness array rather than over it.
The frontal velocity and density structure are measured as the current transitions from flowing along a smooth bed to flowing through the roughness array, and then back to a smooth bed.
We find that, upon entering the roughness array, the gravity current decelerates and the density structure changes from the head and tail structure typical of smooth bed gravity currents, to a wedge shape.
A model is presented that explains the deceleration and change in shape based on a dynamic balance between a pressure gradient within the current tail and a drag force associated with individual roughness elements.
This model accurately predicts the deceleration of the gravity current, supporting the proposed dynamic balance.
\end{abstract}


\maketitle


\section{Introduction}
Gravity currents are flows induced by differences in density between two or more horizontally aligned fluids. 
They are found throughout the environment and are generated both naturally and by human activities. 
Examples of gravity currents include thunderstorm outflows, dust storms, powder-snow avalanches, sea-breeze fronts, gas leaks, and plumes from desalination plants \citep{simpson1999gravity,Ungarish09}.

Due to the importance of gravity currents in a range of engineered and natural situations, they have been extensively studied through numerical, laboratory, and theoretical investigations. 
The majority of these investigations have focused on full-depth lock-exchange gravity currents travelling over smooth boundaries.
The lock-exchange configuration considers the sudden release of a small volume of ``lock'' fluid into a larger ambient fluid and full-depth implies that the lock fluid is vertically homogeneous and of the same depth as the ambient fluid.
Laboratory based examples include \citet{huppert1980slumping}, \citet{LoweIntrusive}, \citet{shin2004gravity}, and \citet{marino2005front} while numerical examples include \citet{hartel2000analysis}, \citet{Birman}, \citet{cantero2007high}, and \citet{ooi2009numerical}. 
In describing such currents, the flow is usually separated into a bulbous head at the front of the current and a slightly less deep tail.

Previous studies have shown that gravity currents develop through several different regimes \citep{Ungarish09}. 
After a short initial transient, a lock-exchange gravity current will enter a slumping regime where the current travels at a constant speed in a quasi-steady state relative to the front. 
In this regime the head of a current mixes with the ambient fluid, predominantly via Kelvin-Helmholtz instabilities. 
The mixed fluid travels backwards with respect to the current head and is  replaced by faster moving, less diluted fluid from the tail \citep{sherLockExchange}.
The replenishment of fluid into the head allows the fluid density within the head to remain constant with time. 
\added{Immediately following the release of the lock fluid, a bore propagates in the opposite direction to the current.
The bore will be reflected by the back wall of the lock and then intersect with the front of the current.
Until the bore nears the front of the current, the current propagates as if the lock were of infinite length.
Following the bore reaching the front of the current, the current decelerates and decreases in size over time, in what is termed the self-similar regime.}
\deleted{Eventually, the remaining lock fluid becomes insufficient to drive the current which causes a bore to propagate along the interface between the current and ambient fluid towards the head. 
Once this bore reaches the head, the current decelerates and decreases in size over time.}
Finally, viscous forces become dominant and the current decelerates more rapidly until it is eventually arrested.
Here we will focus exclusively on the slumping regime.

Motivated by the presence of complex topography in the natural and built environment, several recent studies have investigated the impact of roughness and arrays of bottom-mounted obstacles on gravity currents \citep{peters2000visualization,tokyay2011lock}. 
This work itself, is building on previous studies that examined gravity currents interacting with singular obstructions \citep[e.g.][]{Ermanyuk05,GonzalezJuez09} and complex but continuous bed shapes \citep[e.g.][]{Nicholson15}.
Of particular interest are situations where the roughness or obstacles are of comparable depth to the current itself.
Within the slumping phase, these studies observed slower velocities and enhanced mixing compared to a smooth bed current.
Large eddy simulations (LES) have also been used to investigate lock-exchange gravity currents interacting with fields of square and triangular roughness elements of small scale compared to the current \citep{bhaganagar2017lock}. 
Once again, enhanced mixing was observed and this was linked to the strength of the shear layer formed by the roughness.

Recent studies using laboratory experiments \citep{cenedese2018} and LES \citep{zhou2017propagation} have investigated the impact of a regular array of vertical cylinders on a gravity current.
The current was assumed to be in a quasi-steady state that is equivalent to a smooth bed current in the slumping regime. 
It was found that as the spacing between cylinders decreased, the current transitioned from flowing through the array of cylinders to flowing over the cylinders.
A third regime where the spacing between cylinders was small perpendicular to the flow but large in the flow direction was also investigated. 
In this third regime, the current flows over each row of cylinders individually before plunging to the floor. 
Finally, a fourth regime was identified numerically by \citet{zhou2017propagation} wherein the spacing between cylinders was large perpendicular to the flow and small in the flow direction.
These four regimes were described as through-flow, over-flow, plunging, and skimming, respectively. 
The threshold spacing between regimes was shown to depend on the height of the cylinders relative to the height of the current.
Similar results were found in an inverted study where the cylinders represented roots of floating vegetation \citep{zhang2011exchange} {and for radially spreading gravity currents with both a lock-exchange and continuous source condition \citep{Petrolo22}}.

Very few studies have investigated the transient interactions between gravity currents and fields of roughness. 
\citet{nasr2011turbins} used direct numerical simulations (DNS) to investigate the enhanced mixing and entrainment as well as the three-dimensional vortical structures that are generated by a current interacting with a Gaussian bump and other individual obstacles.
\citet{wilson2017turbulent} observed the increased entrainment and subsequent re-establishment of a current interacting with a rectangular obstacle. 
\citet{kollner2020gravity} employed experiments and DNS to study a gravity current travelling over monodisperse, closely-packed spheres.
It was found that the speed and density of the head reduce with time but that, in contrast to an array of cylinders, the closely packed matrix of the bed slows the rate of dilution into the head. 
To the authors' knowledge no prior studies have investigated the transient behaviour of currents as they transition from a smooth boundary to a rough boundary or from a rough boundary to a smooth boundary. 

The present study employs lock-exchange laboratory experiments to investigate the transient behaviour of a current interacting with a field of vertical cylinders.
We limit our attention to the more commonly studied full-depth gravity currents for ease of comparison with previous work.
The experiments are similar to those of \citet{cenedese2018} but the collection of data over a larger horizontal extent means that we do not need to assume that the currents are in quasi-steady state and can examine the transient behaviour of the currents.
The experimental apparatus and method are described in \S\ref{sec:methods}. \S\ref{sec:results} describes the experimental results with particular emphasis on the frontal velocity, the density structure, and how these quantities develop in space and in time. 
Based on the experimental results, an analytical model is presented in \S\ref{sec:Model} before final conclusions are provided in \S\ref{sec:conclusion}.

\section{Methodology}
\label{sec:methods}
\begin{figure}
    \centering
    \includegraphics[width=0.9\linewidth]{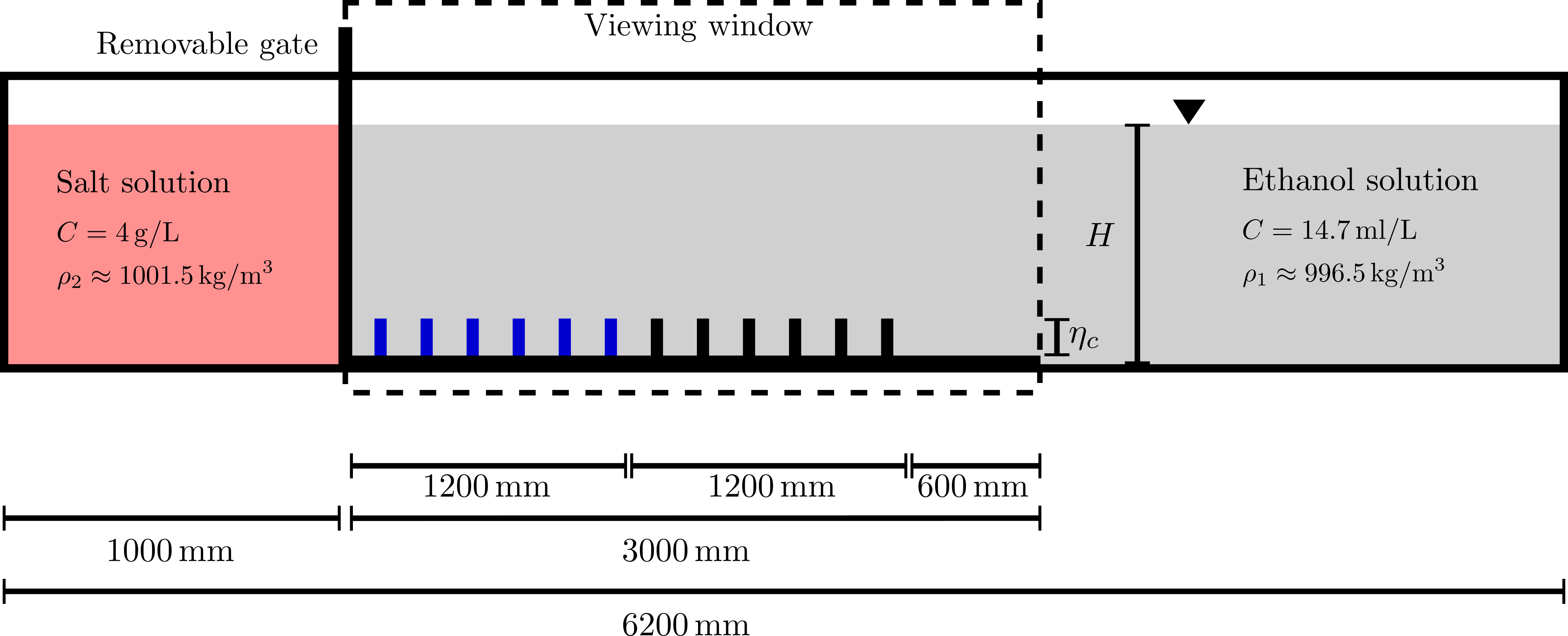}
    \caption{Schematic of the experimental apparatus. Initially, dense fluid in the lock on the left is separated from lighter, ambient fluid on the right by a gate. Upon removal of the gate, a dense gravity current forms that either immediately interacts with the roughness or develops as a smooth bed current before encountering the roughness. 
    \replaced{Roughness elements marked in blue were only used for the `gate' experiments and roughness elements marked in black were only used for the standard configuration experiments.} {In the schematic, all possible roughness elements are shown but in a given experiment, roughness elements only occuppied 1.2\,m of the tank.}
    \added{Four cameras observe the flow across the viewing window marked by dashed lines.}
    \deleted{The flow is recorded by four cameras placed along the channel. Not to scale.}}
    \label{fig:expSchematic}
\end{figure}

Experiments were conducted in a 6.2\,m long and 0.25\,m wide Perspex channel as shown in Fig.~\ref{fig:expSchematic}. 
The left end consisted of a 1\,m long lock that contained dense fluid.
The lock fluid was initially separated from the lighter, ambient fluid to the right by an approximately 1\,mm thick stainless steel gate. 
To the right of the lock was a 3\,m long false floor with holes present to screw roughness elements into the bed. 

\begin{figure}
    \centering
    \includegraphics[width=0.9\linewidth]{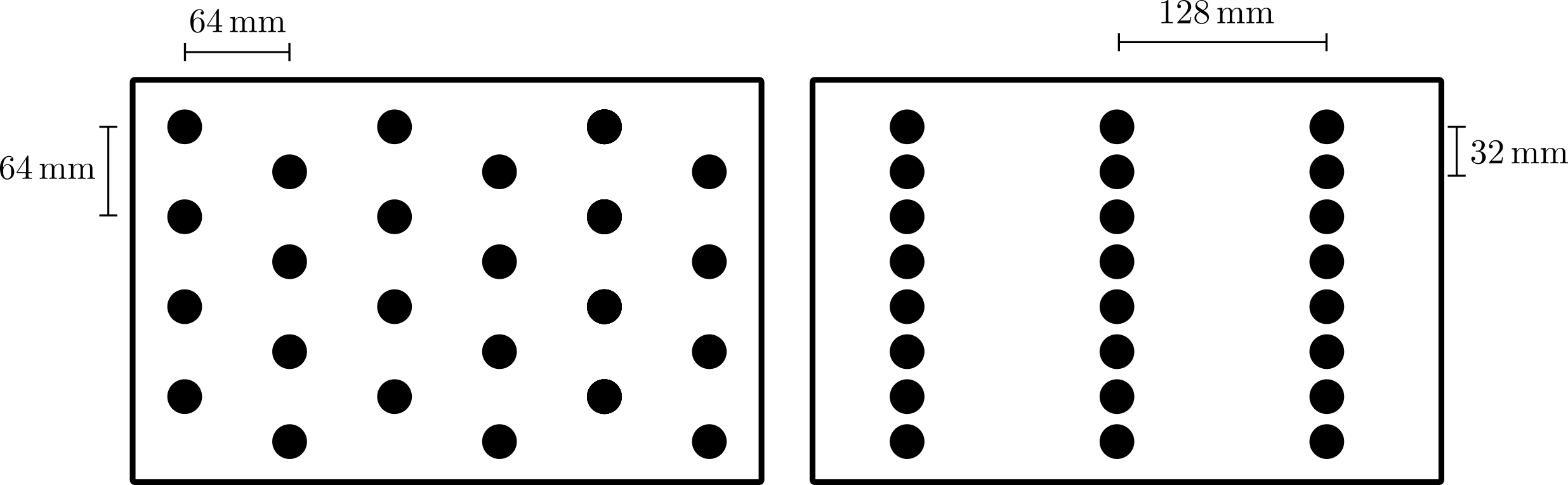}
    \caption{The two roughness configurations used in the experiments. The sparse configuration is shown on the left panel and the plunging configuration on the right panel. Not to scale.}
    \label{fig:roughnessConfiguration}
\end{figure}

Two roughness configurations were investigated as shown in the plan-view schematic in Fig.~\ref{fig:roughnessConfiguration}.
In both cases the roughness was formed from a regular array of  cylinders with diameters of 20\,mm and heights of 50\,mm.
The first roughness configuration is similar to the sparse configuration investigated by \cite{cenedese2018} and consisted of an equally spaced, staggered array of cylinders with a distance between cylinder centers of 64\,mm. 
The second roughness configuration had distances between cylinder centers of 32\,mm across the channel and 128\,mm in the flow direction.
Following the analysis presented in \cite{zhou2017propagation}, these two roughness configurations are referred to as the sparse and plunging configuration, respectively.
Whenever holes in the false floor were not occupied with cylinders, they were covered with tape to form a smooth bed.
We note that the fraction of the tank base occupied by cylinders was equal for the two configurations; the only difference was the arrangement of the cylinders.

\added{For most of the experiments, only the roughness elements shown in black in Fig.~\ref{fig:expSchematic} were present.
The roughness elements shown in blue were removed such that the first 1.2\,m of the false floor had a smooth bed.
This configuration allowed the current to establish before encountering a 1.2\,m long section of roughness elements.
The current was observed continuously from the gate until 0.6\,m after the roughness array, as shown by the dashed lines in Fig.~\ref{fig:expSchematic}.}
\deleted{Although the roughness elements are shown along the entire 3.0\,m false floor in Fig.~\ref{fig:expSchematic}, for most experiments the first 1.2\,m of the false floor was left without roughness elements to allow the current to establish before encountering a 1.2\,m long section of roughness.}
\added{A second set of experiments, referred to as the gate experiments, was conducted where only the roughness elements shown in blue were present. 
As in the standard experiments, the current was observed for 3\,m after the gate in these experiments.
This allowed the current to be observed through the roughness array, and then for a further 1.8\,m, as the current transitions back to a smooth bed.}
\deleted{A second set of experiments, referred to as the `gate' experiments, was conducted where the roughness elements were present immediately from the gate.
For these gate experiments, the length of the roughness was maintained at 1.2\,m but the current was observed for a further 1.8\,m beyond the roughness.}
The gate experiments were only performed in the sparse configuration and fulfilled two purposes.
First, they allowed for a comparison between currents that were and were not allowed to develop before encountering roughness.
Second, they allowed a more complete investigation of the behaviour as currents emerged from a field of roughness onto a smooth boundary.


The fluid initially in the lock was a 4\,g/l NaCl solution with a nominal density of $\rho_2=1001.5\,{\rm kg/m^3}$.
The ambient fluid was a 14.7\,ml/l denatured ethanol solution with a nominal density of $\rho_1=996.5\,{\rm kg/m^3}$.
A denatured ethanol solution was used in the ambient fluid to ensure that the two fluids had equal refractive indices.
The exact fluid densities were measured using an Anton Parr density meter with an accuracy of $0.004\,{\rm kg/m^3}$ and the density differences can be found in table~\ref{tab:experiments}.
For each roughness configuration experiments were run with nominal fluid depths of 100, 150, 200, and 270\,mm. 
The exact fluid depths were measured before each experiment and can be found in table~\ref{tab:experiments} with other important experimental parameters.
{Due to the different fluid heights in each experiment, the time and distance before the bore from the lock release reached the head of the current varied between experiments.
To ensure that this didn't affect the results, all measurements were taken before the lock bore reached the current head.}


Four JAI GO-5101C-PGE cameras with zoom lenses were \added{positioned approximately 5\,m from the front of the tank and were }used to capture the flow.
\deleted{Approximate locations of these cameras are shown in Fig.~\ref{fig:expSchematic}.} 
For the standard roughness configurations, the cameras observed the current from the gate, across the full roughness array, and for a further 0.6\,m beyond the roughness array.
Each camera captured images with resolutions of 2464 x 2056 pixels and a frame rate of 22.7 Hz. 
Experiments were all repeated twice; once to measure width-averaged density using Light Attenuation (LA) and once to measure centerline velocities using Particle Tracking Velocimetry (PTV).
PTV results are not used in a quantitative way within this paper so are {only briefly} described here.
However, the measurement system is {fully }described within \citet{kollner2020gravity} for related experiments.
Several experiments were repeated multiple times to assess the repeatability of the measurements.
Typical uncertainties for the derived parameters (defined in \S\ref{sec:ND}) did not exceed 3\%.
For both measurement techniques, the analysis was performed in the software \textit{Streams} \citep{streams} and the results were interpolated onto rectangular, Eulerian grids with 4\,mm resolution in the flow direction and 2\,mm resolution in the vertical direction.

For the light attenuation experiments, a small amount of red carmoisine dye was added to the lock fluid.
The tank was backlit by a bank of LED lights that were placed behind a diffuser sheet to ensure even lighting.
{The light attenuation system was regularly calibrated throughout the experimental program such that  fluid densities could be inferred from the light intensity observed by the cameras.
The calibration was conducted by measuring the optical thickness of the tank at a range of dye concentrations, where the optical thickness is defined as $d = \ln(I_0/I)$, $ I_0 $ is the light intensity when the tank contains undyed water and $ I $ is the observed light intensity.
Measurements of the optical thickness were made using seven different dilutions of the lock fluid that varied in concentration from fresh water to 120\% of the lock concentration.
The density of these solutions was measured using an Anton Parr density meter.
Finally, a linear fit between optical thickness and fluid density was made on a pixel-by-pixel basis that allowed for the calculation of fluid densities throughout an experiment.}

{For the PTV experiments, both the ambient fluid and the lock fluid were seeded with particles of pliolite resin with diameters ranging from $ 180-250\,{\rm \mu m} $.
The particles were illuminated by a 10\,mm wide light sheet generated by an array of LED lights placed above the tank.
The light sheet was positioned along the flow direction and at the center of the tank.
Particles were recorded using the same cameras as for the light attenuation experiments.}

A full list of experiments is found in table~\ref{tab:experiments}.
The table provides the measured fluid depth, $H$, the non-dimensional cylinder height, $\sigma=\eta/H$, the density difference between the lock and the ambient fluid, $\Delta \rho$, the calculated reduced gravity, $g'=g\Delta\rho/\rho_2$, the buoyant velocity, $u_B=\sqrt{g'H}$, and the Reynolds number, ${\rm Re}_B=u_BH/\nu$.
In the above quantities, $g$ is the acceleration due to gravity and $\nu=10^{-6}\,{\rm m^2/s}$ is the kinematic viscosity of the fluid.
The experiments are labelled based on the roughness configuration (``S" for sparse or ``P" for plunging) and the non-dimensional cylinder height.
A ``G" is appended to the name for the `gate' experiments described above.
As an example, experiment S49G had cylinders arranged in the sparse configuration, a non-dimensional cylinder height of 0.49, and cylinders present from the gate.





\begin{table}
\caption{\label{tab:experiments}
Fluid depth, $H$, non-dimensional roughness height, $\sigma$, density difference, $\Delta\rho$, reduced gravity, $g'$, buoyant velocity, $u_B$, and Reynolds number ${\rm Re}_B$ for all experiments. $H$ and $\Delta\rho$ are directly measured while all other quantities are calculated.}
\begin{ruledtabular}
\begin{tabular}{lcccccc}
Name & $H$ & $\sigma$   & $\Delta{\rho}$   & $g'$      & $u_B$  & ${\rm Re}_B$\\
    &   (m) &           & ($\rm{kg/m^3}$) & (m/s$^2$)  & (m/s)  &       \\
\hline
S48 &  0.104 & 0.48 & 5.61 & 0.0549 & 0.076 & 7900  \\ 
S32 &  0.154 & 0.32 & 5.62 & 0.0550 & 0.092 & 14200 \\
S25 &  0.202 & 0.25 & 5.59 & 0.0548 & 0.105 & 21200 \\
S18 &  0.278 & 0.18 & 5.59 & 0.0548 & 0.123 & 34300 \\
S49G & 0.102 & 0.49 & 5.03 & 0.0490 & 0.071 & 7200  \\
S33G & 0.150 & 0.33 & 5.08 & 0.0500 & 0.086 & 13000 \\
S25G & 0.201 & 0.25 & 4.60 & 0.0450 & 0.095 & 19100 \\
S19G & 0.270 & 0.19 & 5.08 & 0.0500 & 0.116 & 31400 \\
P50 &  0.100 & 0.50 & 4.80 & 0.0470 & 0.069 & 6900  \\ 
P32 &  0.154 & 0.32 & 5.76 & 0.0564 & 0.093 & 14300 \\ 
P25 &  0.204 & 0.25 & 5.59 & 0.0548 & 0.106 & 21600 \\ 
P18 &  0.274 & 0.18 & 5.73 & 0.0561 & 0.124 & 34000 \\ 
\end{tabular}
\end{ruledtabular}
\end{table}

\section{Results}
\label{sec:results}
\subsection{Non-dimensionalisation and analysis framework}
\label{sec:ND}
For the remainder of this paper, all quantities are non-dimensional unless otherwise stated.
Spatial quantities have been non-dimensionalised by the fluid depth, $H$, velocities have been non-dimensionalised by the buoyant velocity, $u_B$, times have been non-dimensionalised by $T_B = H/u_B$, and fluid densities have been non-dimensionalised such that undiluted lock fluid has a density of 1 and undiluted ambient fluid has a density of 0. 

The origin of the spatial coordinates is defined to be the {location of the lock gate} in the horizontal ($x$) direction and the bottom of the tank in the vertical ($y$) direction.
The temporal origin is defined to be {the time when the lock gate is removed and the current is initiated}.
{We additionally define $ x_1 $ and $ x_2 $ as the front face and the end of the roughness array, respectively.
Similarly, $ t_1 $ and $ t_2 $ refer to the times when the current front reaches $ x_1 $ and $ x_2 $.
Since the roughness array is at a fixed location, the non-dimensional location of the roughness, $x_1$ to $x_2$, varies with the fluid depth (or, equivalently, the non-dimensional roughness height).
This is unlikely to affect the flow development as discussed in Sec.~\ref{sec:Qualitative} and to simplify the analysis, results are typically presented in terms of $\hat{x}=(x - x_1)$ and $\hat{t} = (t-t_1)$.}

Following previous work \citep{shin2004gravity}, we define the {buoyant height} of the current as
\begin{equation}
h_B(x,t)  =  \int_{0}^1 \rho\, {\rm d}y.
\end{equation}
The buoyant height is indicative of the driving force within the current and depends on both the density of the current and its vertical extent.
At times it can be useful to separate these two components.
For this purpose, we also define a {current envelope}, $h_c(x,t)$, that isolates the vertical extent of the current. 
The current envelope is defined as the maximum $y$ location where the measured density is greater or equal to 0.05.

The front location $x_F(t)$ is defined as the farthest downstream location at which $h_B(x,t) \geq 0.02$.
Using a slightly different threshold for the buoyant height does not significantly affect the front location.
The front velocity of the current is then given by the derivative of $x_F(t)$ with respect to time.
In our non-dimensional system, the front velocity is equivalent to the Froude number and will be referred to as such throughout.

{We note here that there are numerous non-dimensional quantities that describe the geometry of the roughness elements and the roughness field.
With the exception of the non-dimensional cylinder height, $ \sigma $, we are keeping these constant, but several other parameters have been explored in earlier work \citep{cenedese2018,zhou2017propagation}.
Of particular note is the roughness element aspect ratio and the plan density ($\alpha$ and $ \sigma $ in the nomenclature of \citep{cenedese2018}).
The roughness element aspect ratio has been shown to have an insignificant effect on the propagation of a gravity current while variations in the plan density are beyond the scope of this study.}


\subsection{Qualitative description of the gravity current development}
\label{sec:Qualitative}
\begin{figure}
    \centering
    \includegraphics[width=0.96\linewidth]{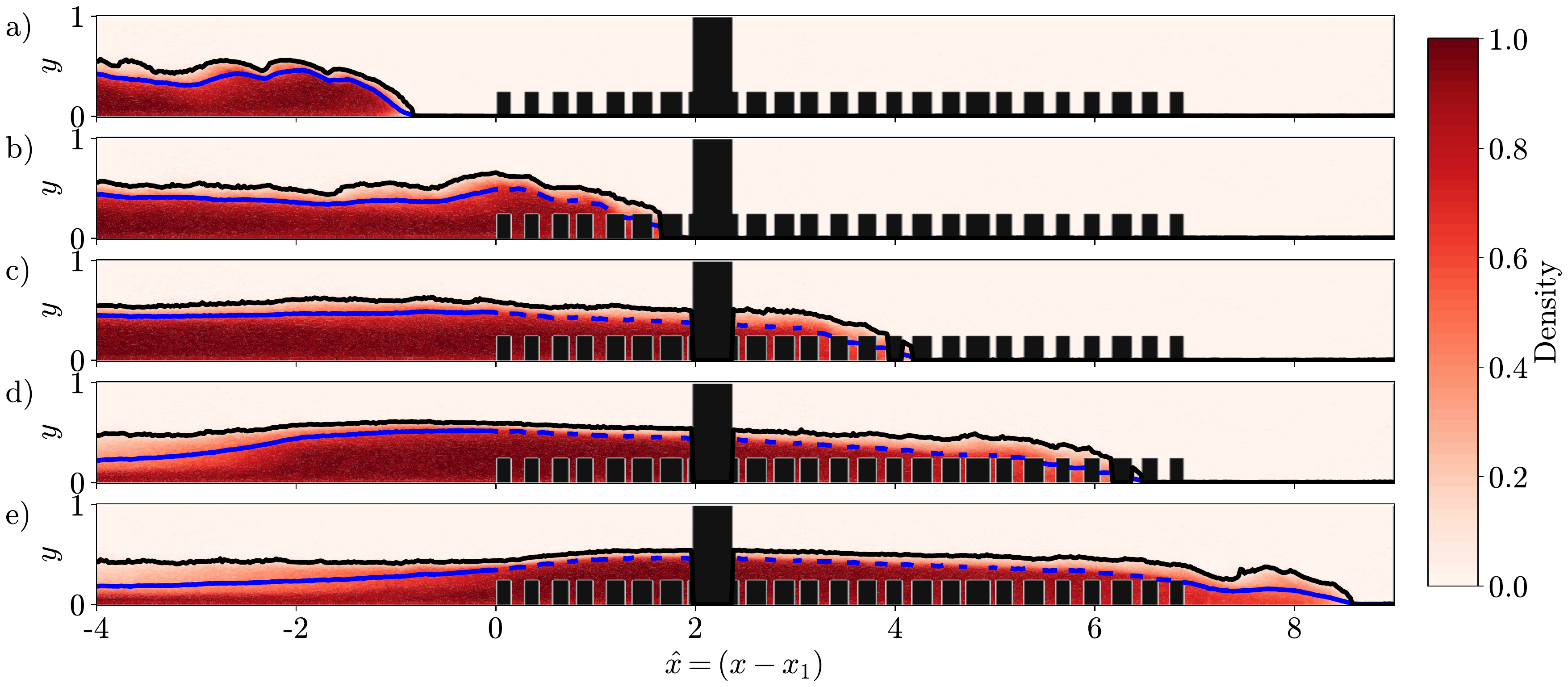}
            \caption[Instantaneous density field for S5H20 experiment at different times]{Width-averaged density field for experiment S25 at times of $\hat{t}=-2$, 4, 10, 16 and 22. The black line shows the current envelope, $h_c$, and the blue line shows the buoyant height, $h_B$. Black cylinders represent locations where the flow was obscured from the camera, either due to roughness elements or, at $\hat{x}\approx 2$, a vertical brace on the side of the tank.}
            \label{fig:S5H20}
\end{figure}

Fig.~\ref{fig:S5H20} shows the instantaneous density field for experiment S25 at five different times. 
Black rectangles correspond to locations where visualisation was unavailable, either due to roughness elements or, at $\hat{x}\approx 2$, a full-height supporting brace on the outside the tank. 
The roughness elements appear to be slightly different in size due to the data being interpolated onto a regular grid that doesn't align with the obstacle locations.
\added{This has no effect on the measurement values but limits the locations where some measurements can be made (e.g. $ h_B $ can only be measured where the fluid density is available from the bed to the free surface).}
Before the current reached the roughness (Fig.~\ref{fig:S5H20}(a)) the current is observed to have fully developed in the slumping phase with Kelvin-Helmholtz instabilities present in the shear layer between the two fluids. 
At this time $h_B$ and $h_c$ show a head-and-tail structure that is consistent with previous observations of smooth bed gravity currents \citep{huppert1980slumping}. 

When the current reached the roughness array, fluid was deflected upwards and the buoyant height and current height became larger than that of  a smooth bed current.
While the current propagated through the roughness, a bore, hereafter referred to as the roughness bore, was reflected back upstream. 
The roughness bore was reflected from the front of the roughness and travelled upstream until it encountered the bore that was reflected from the end wall of the lock.

As the current travelled through the roughness, it transitioned from having a head and tail structure to forming a wedge-like structure. 
{The wedge structure is best understood by examining the buoyant height profile behind the current, as this shows the driving forces within the current.
This will be analysed for both the sparse and the plunging roughness configurations more quantitatively in later sections.}
The location of maximum height (considering either $h_c$ or $h_b$) ceased to be immediately behind the front of the current and instead became stationary at the upstream end of the roughness (Fig.~\ref{fig:S5H20}(c) and~(d).
The wedge shape, seen when  the nose reached the final row of cylinders, is consistent with the through-flow regime described in \cite{cenedese2018}.
The wedge shape is similar to that observed in flows with full-depth cylinders \citep{tanino2005aquatic} where currents developed to a wedge shape as they transitioned from an inertia dominated to a drag dominated regime.
Beyond the shape of the current, Fig.~\ref{fig:S5H20} also shows that the density at the front decreased as the current travelled, unlike smooth bed currents where replenishment of dense fluid from the tail results in a constant fluid density in the head \citep{sherLockExchange}.

\begin{figure}
    \centering
        \includegraphics[width=0.96\linewidth]{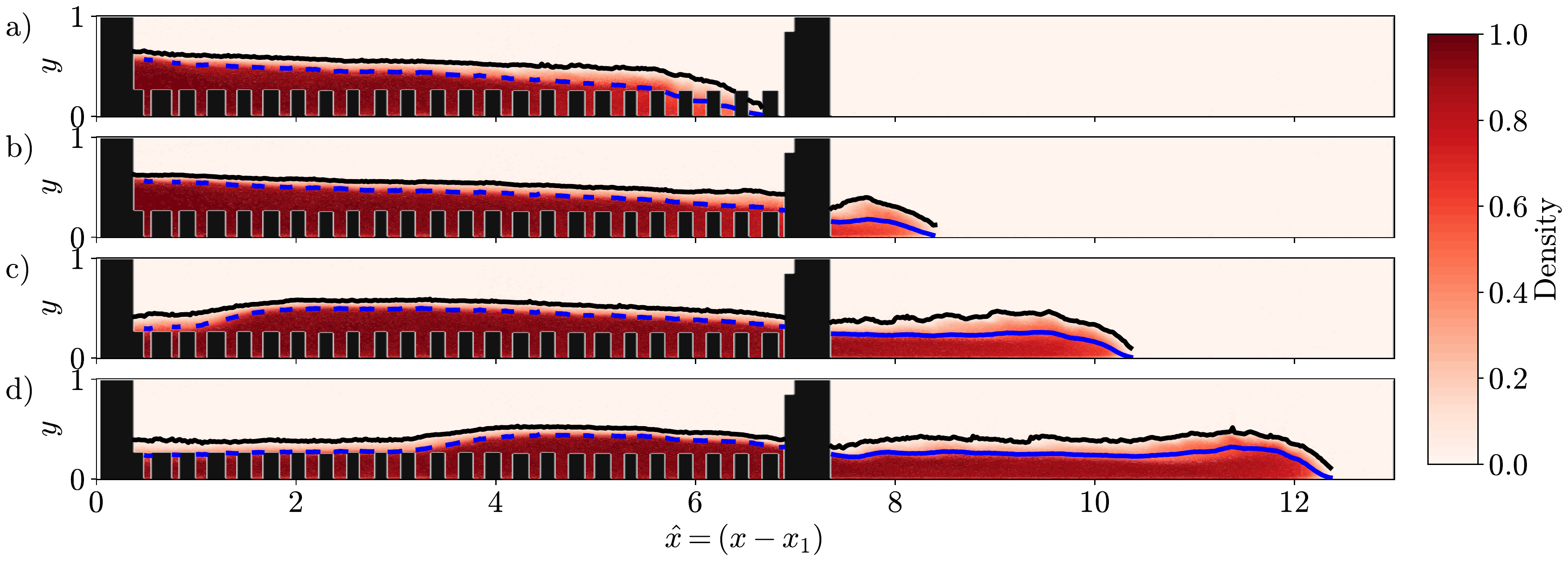}
            \caption[Instantaneous density field for S25G experiment at different times]{Width-averaged density field for experiment S25G at times of $\hat{t}=16$, 21, 26 and 31. The black line shows the current envelope, $h_c$, and the blue line shows the buoyant height, $h_B$. Black cylinders represent locations where the flow was obscured from the camera, either due to roughness elements or, at $\hat{x}\approx 0$ and 7, vertical braces on the side of the tank.}
            \label{fig:S5H20E}
\end{figure}

Fig.~\ref{fig:S5H20}(e) shows the current after it has exited the roughness and re-established as a smooth bed current with a typical head-and-tail structure. 
To better explore this re-establishment, Fig.~\ref{fig:S5H20E} shows an instantaneous density field for the experiment S25G at four different times. 
Fig.~\ref{fig:S5H20E}(a) is from the same time as Fig.~\ref{fig:S5H20}(d) and hence provides a direct comparison between the two experiments. 
{Currents were both qualitatively and quantitatively (see later sections) similar regardless on whether the roughness array started downstream or immediately at the gate, suggesting that whether or not a current is allowed to develop before encountering roughness has a minimal impact on the final state of the current.
An important implication of this similarity for our experiments is that the different non-dimensional distances that currents travelled before reaching the roughness array will not significantly affect the results.}

As the current shown in Fig.~\ref{fig:S5H20E} emerged from the roughness, it recovered the head and tail structure typical of smooth bed currents. 
The head grew in vertical extent and its density increased. 
This behaviour suggests that although the head replenishment process described in \cite{sherLockExchange} was diminished within the roughness array, it rapidly restarted once the current emerged from the roughness{ array}.
Equivalently, dilution of the head fluid is more rapid than replenishment of dense fluid from the tail while the current is within the roughness array but less rapid than replenishment immediately after the current emerges from the roughness. 
It remains unclear from the qualitative results whether this behaviour is due to changes in the dilution rate, the replenishment rate, or a combination of both.

\begin{figure}
	\centering
	\includegraphics[width=0.96\linewidth]{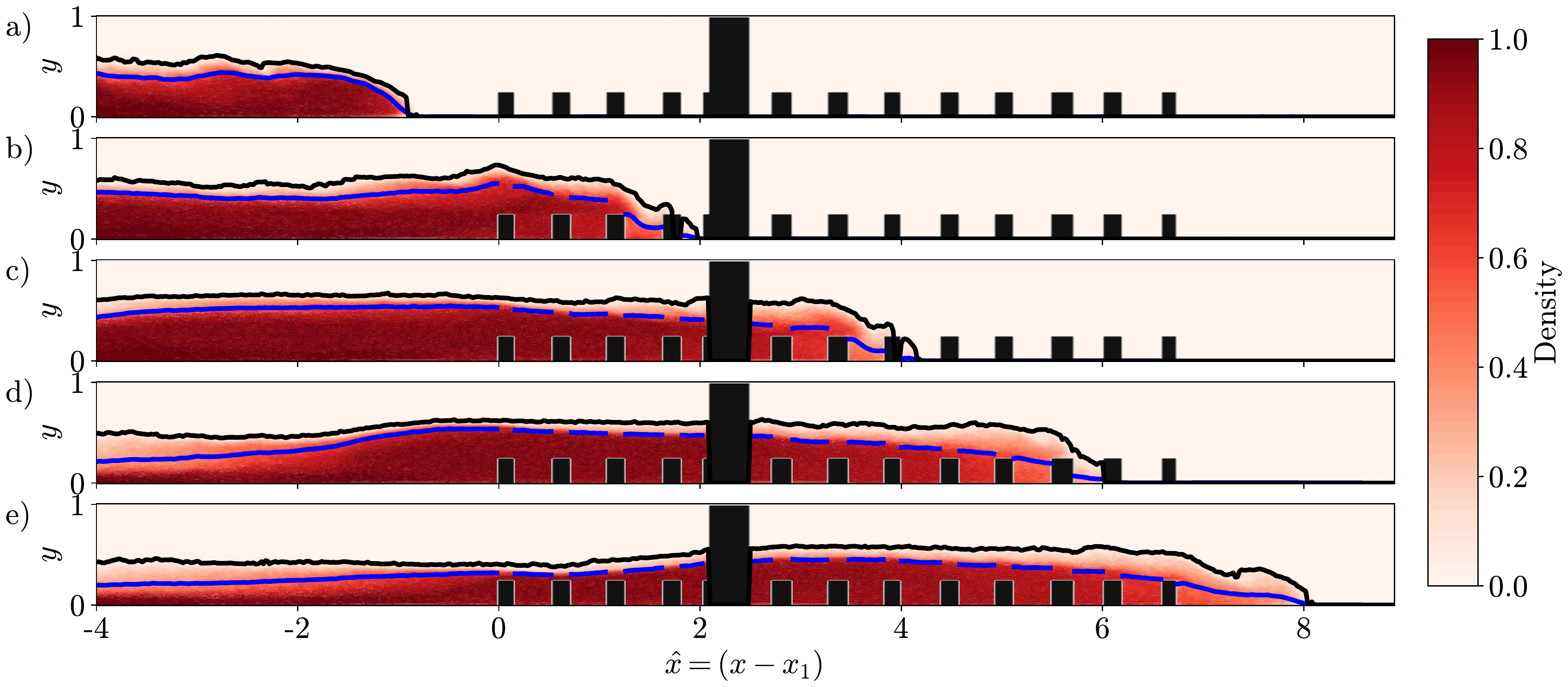}
	\caption{\added{Width-averaged density field for experiment P25 at times of $\hat{t}=-2$, 5, 12, 19 and 26. The black line shows the current envelope, $h_c$, and the blue line shows the buoyant height, $h_B$. Black cylinders represent locations where the flow was obscured from the camera, either due to roughness elements or, at $\hat{x}\approx 2$, a vertical brace on the side of the tank.}}
	\label{fig:P5H20}
\end{figure}

Figs.~\ref{fig:S5H20} and~\ref{fig:S5H20E} are representative of all of the sparse configuration experiments.
In all cases the currents had the through-flowing phenomenology \citep{cenedese2018} and formed a wedge structure within the roughness before returning to a head-and-tail structure after the roughness.
The plunging configuration experiments also tended to form a wedge structure as they travelled through the roughness.
\deleted{However, the mechanisms and flow paths by which they travelled were notably different.}
\added{For example, Fig.~\ref{fig:P5H20} shows the instantaneous density field for  the P25 experiment at five different times. 
As in earlier figures, the black line shows the current envelope and the blue line shows the buoyant height.
When considering the buoyant height profile as representative of the structure of the current, a wedge structure is observed through the tail of the current within the roughness array (e.g., Fig.~\ref{fig:P5H20}d).}

\begin{figure}
	\centering
	\includegraphics[width=0.6\linewidth]{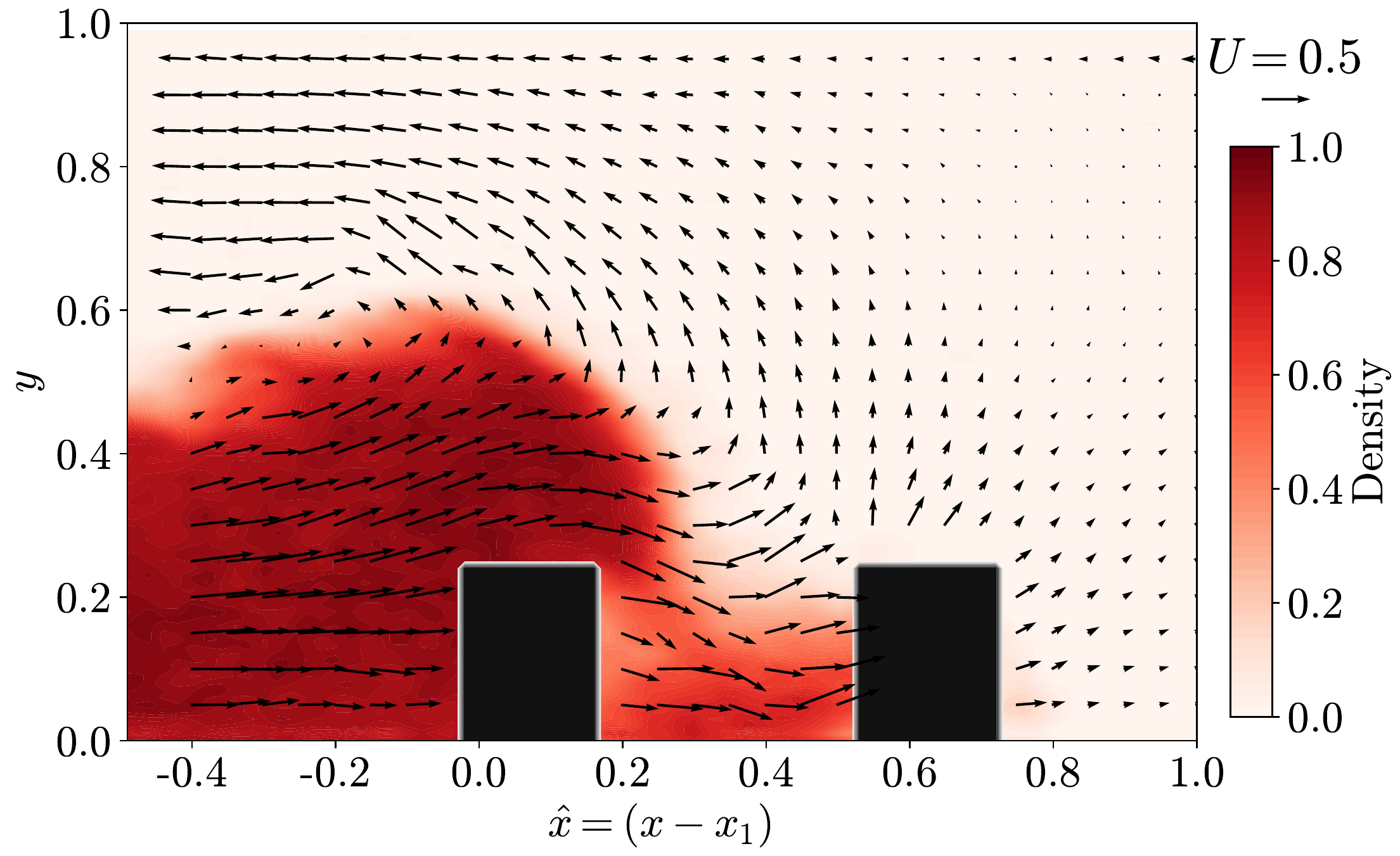}
	\caption[Instantaneous density field for P25 experiment shortly after encountering the first row of roughness]{Width-average density and centerline velocity measurements for experiment P25 as the current nose reached the second row of roughness. False colour shows fluid density as measured by LA and arrows show velocity vectors as measured by PTV. Note that the two fields are from different realisations of the same experiment.}
	\label{fig:P5H20Plunging}
\end{figure}

\added{Despite the sparse and plunging configuration experiments both forming a wedge structure, the flow paths that fluid travelled were notably different in the plunging configuration.}
When the roughness height was sufficiently large, currents would plunge individually over each row of roughness.
An example is shown in Fig.~\ref{fig:P5H20Plunging} for experiment P25.
Fig.~\ref{fig:P5H20Plunging} shows the density field (colour) and the velocity field (arrows) from two realisations of the same experiment overlain upon one another.
For both realisations, a time shortly after the current has reached the first row of roughness is shown.
A small volume of fluid propagates through the gaps in the cylinder array and is seen in front of the rest of the current ($\hat{x}\approx0.5$).
However, the majority of the fluid is deflected over the cylinders and plunges downwards before encountering the following row of cylinders.
During this plunging process, a small region of light, ambient fluid is trapped underneath the current just downstream of the cylinder.
The trapped ambient fluid will mix with the fluid in the current.
Depending on the rate of mixing, the ambient fluid will either become fully mixed into the current or travel vertically through the current and intrude below the overlying ambient fluid.

\begin{figure}
    \centering
    \includegraphics[width=0.95\linewidth]{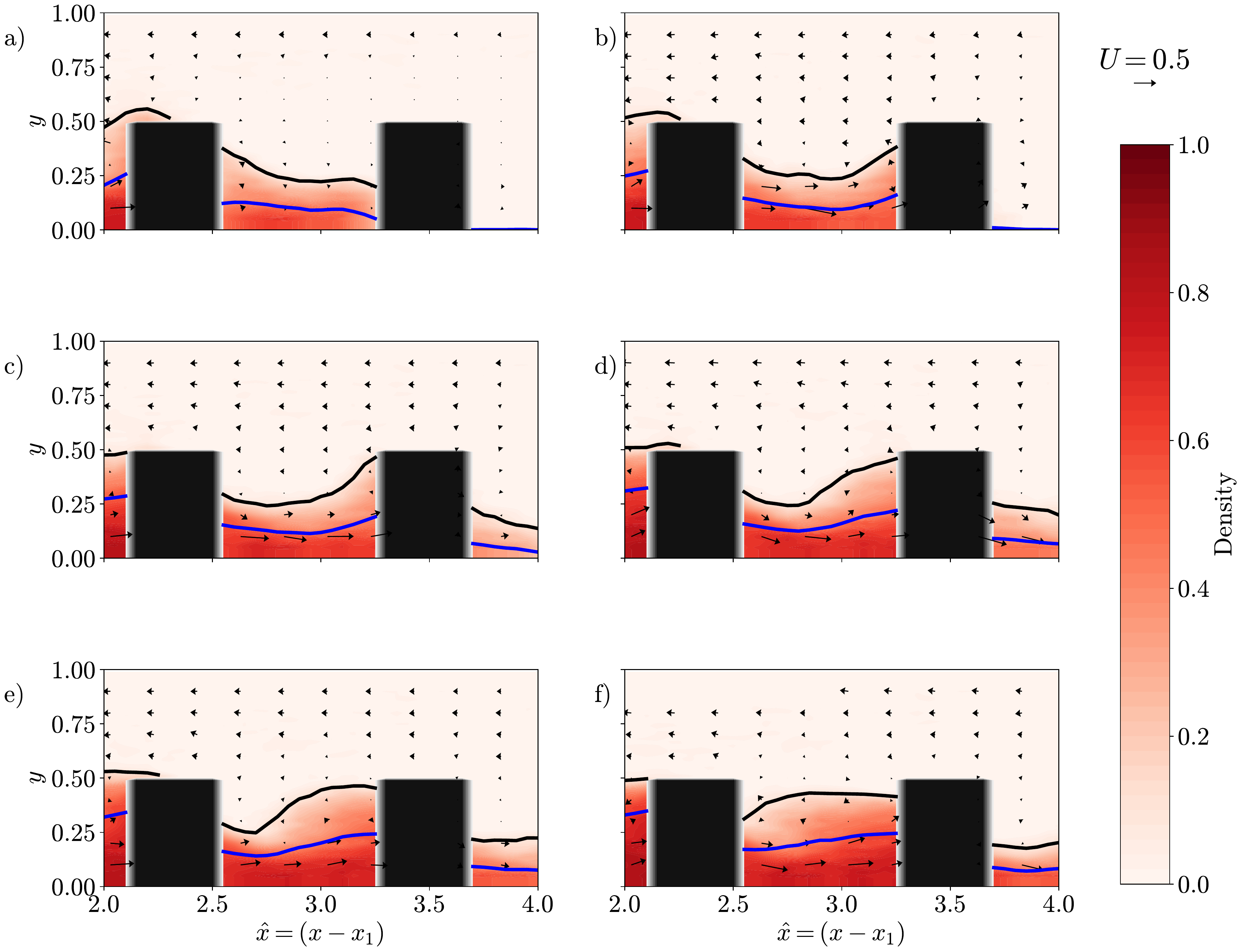}
    \caption[Instantaneous density fields showing bores between rows of cylinders for P50 experiment]{Width averaged density fields and velocity vectors between the third and fourth cylinders for the P50 experiment at $\hat{t} = 8$, 9, 10, 11, 12 and 13. False colour shows fluid density as measured by LA and arrows show velocity vectors as measured by PTV. Note that the two fields are from different realisations of the same experiment. The black line shows the current envelope, $h_c$, and the blue line shows the buoyant height, $h_B$.}
    \label{fig:PlungingBores}
\end{figure}

When the roughness height is larger, the current is no longer deflected above the roughness height and is unable to plunge over the roughness.
This behaviour is shown in Fig.~\ref{fig:PlungingBores} for experiment P50.
As the current reaches a given row of roughness, some fluid immediately propagates through the gaps between roughness elements and some is deflected vertically. 
However, the vertical deflection never exceeds the height of the roughness and all fluid eventually propagates through the gaps between roughness elements.
{It is possible that such an effect could be caused by the ambient return flow constraining how high the current can travel.
However, the return flow is much weaker (relative to the current velocity) in the P50 experiment than in the P25 experiment, suggesting that it is not driving the observed difference in dynamics.}
As such, the dynamics more closely resemble the sparse configuration behaviour than the plunging behaviour seen in Fig.~\ref{fig:P5H20Plunging} involving large vertical velocities and high levels of mixing behind each row of roughness.

\subsection{Current velocity}

\begin{figure}
\centering
\begin{tabular}{cc}
\includegraphics[width=0.48\textwidth]{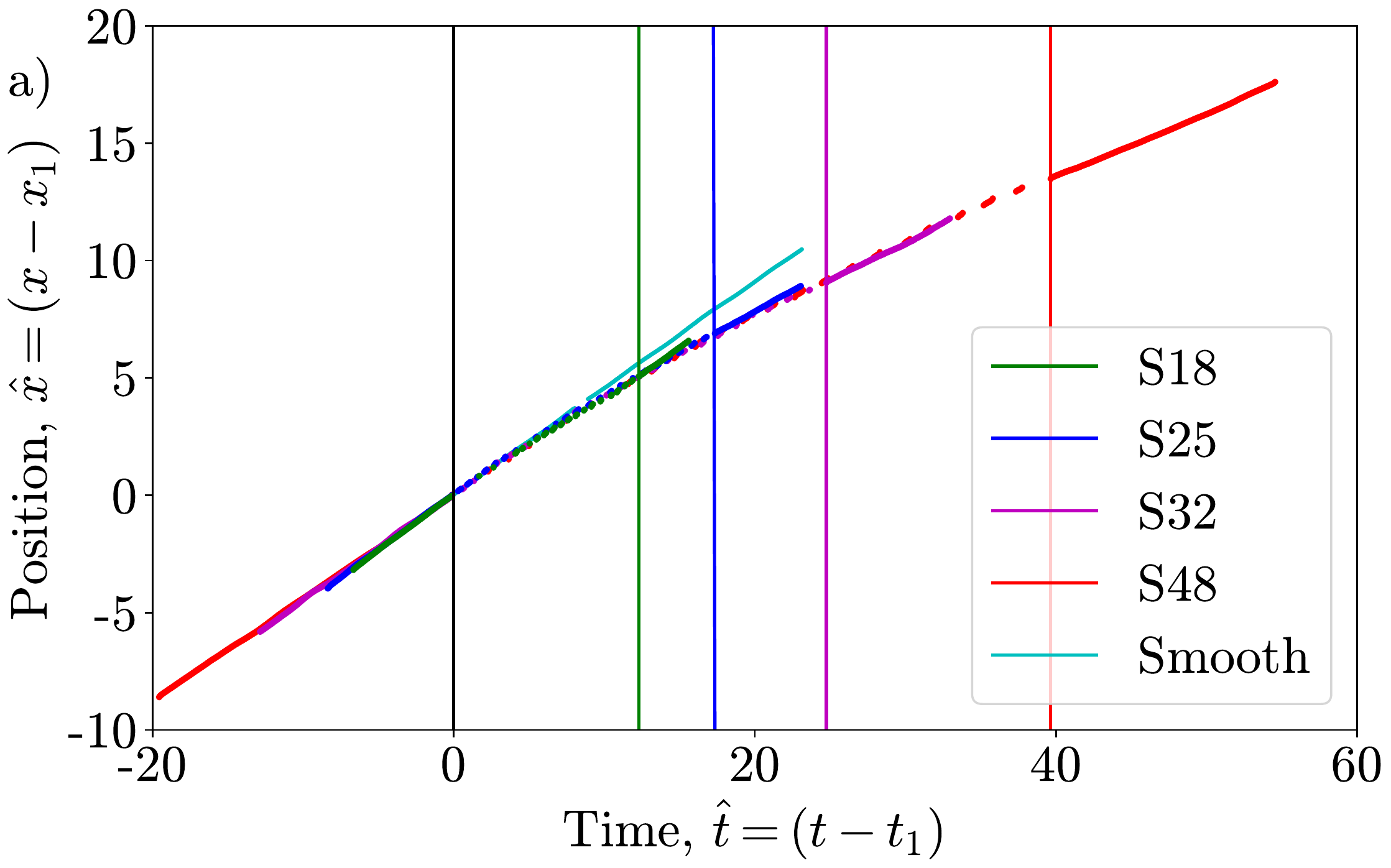}&
\includegraphics[width=0.48\textwidth]{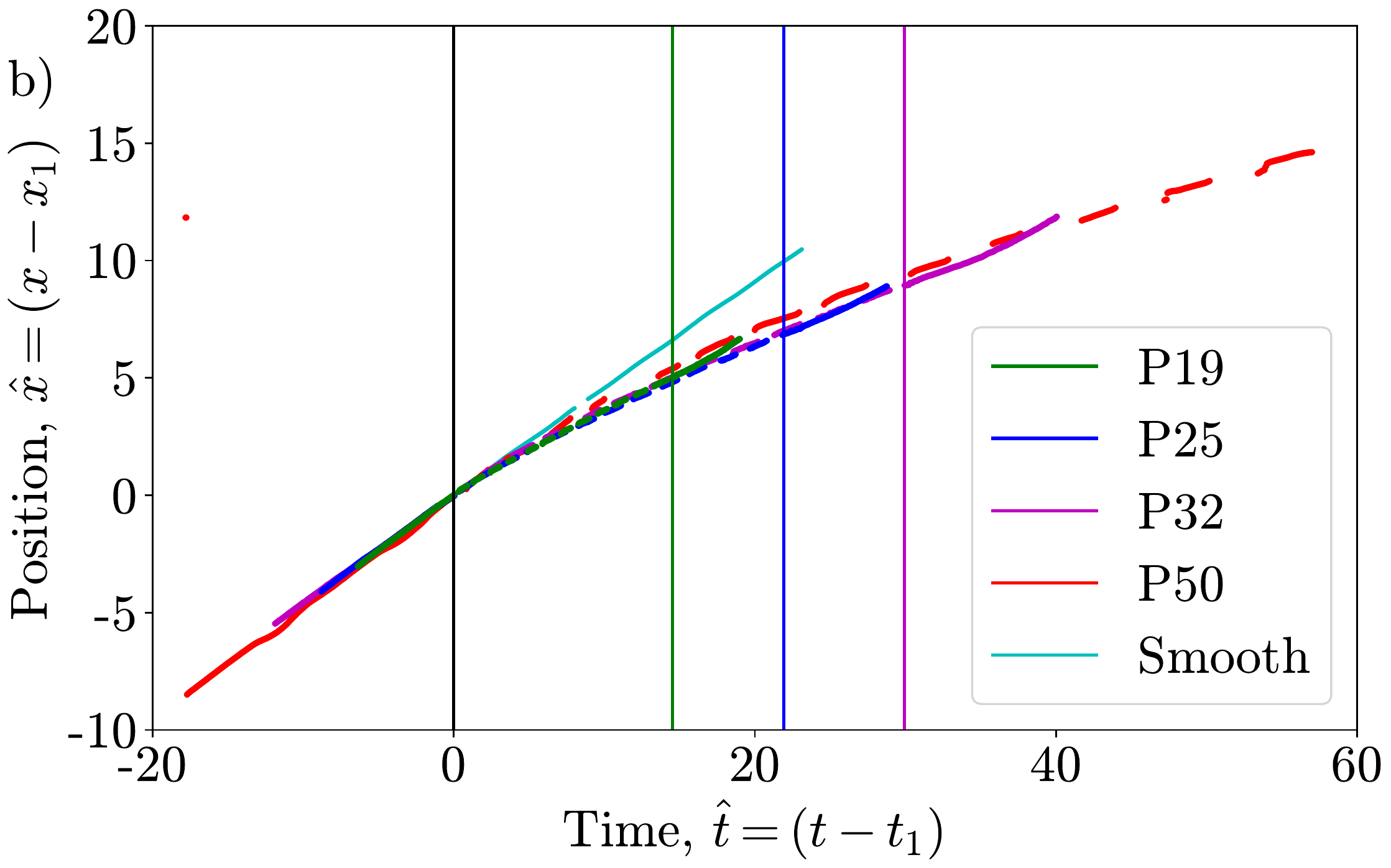}\\
\includegraphics[width=0.48\textwidth]{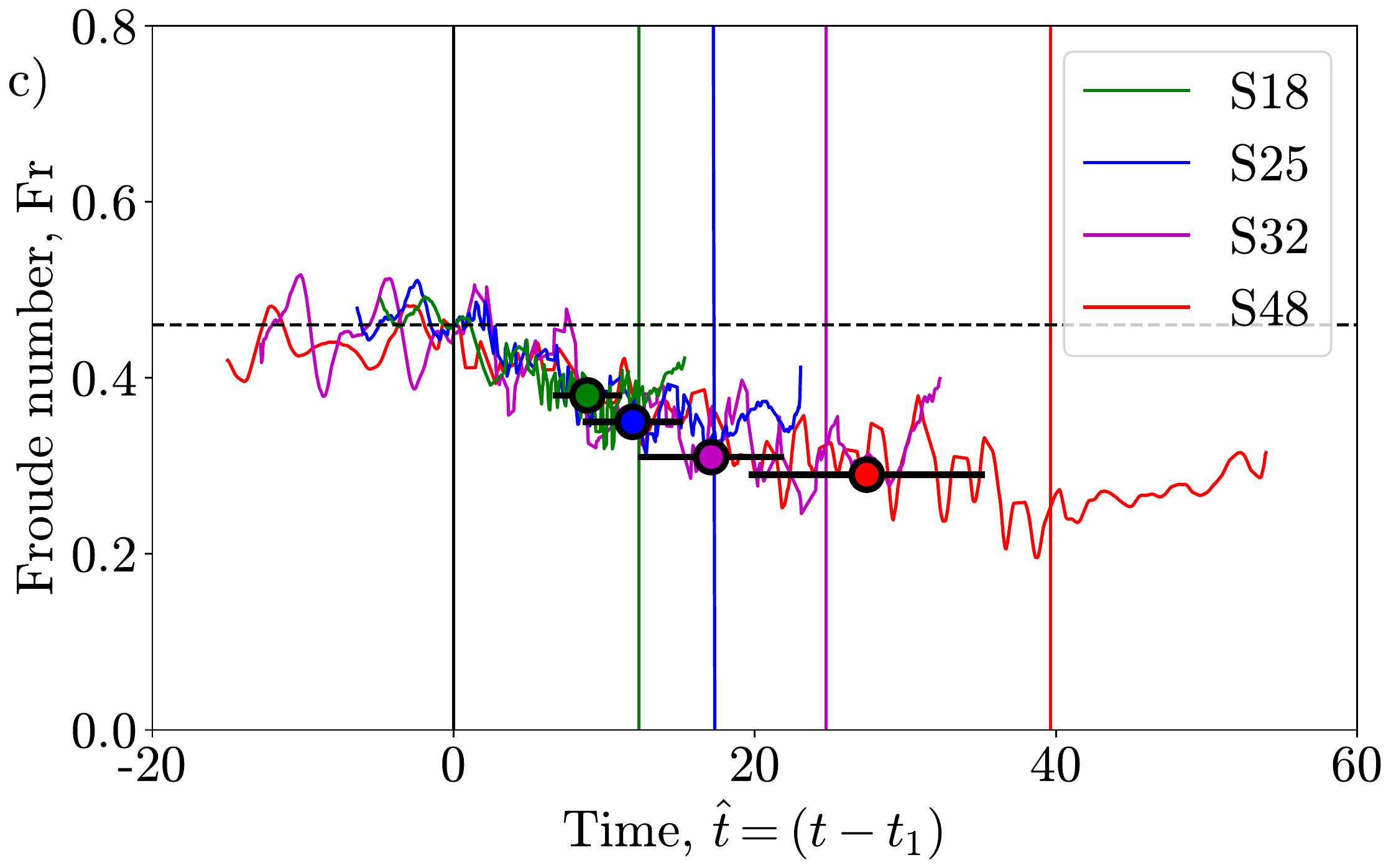}&
\includegraphics[width=0.48\textwidth]{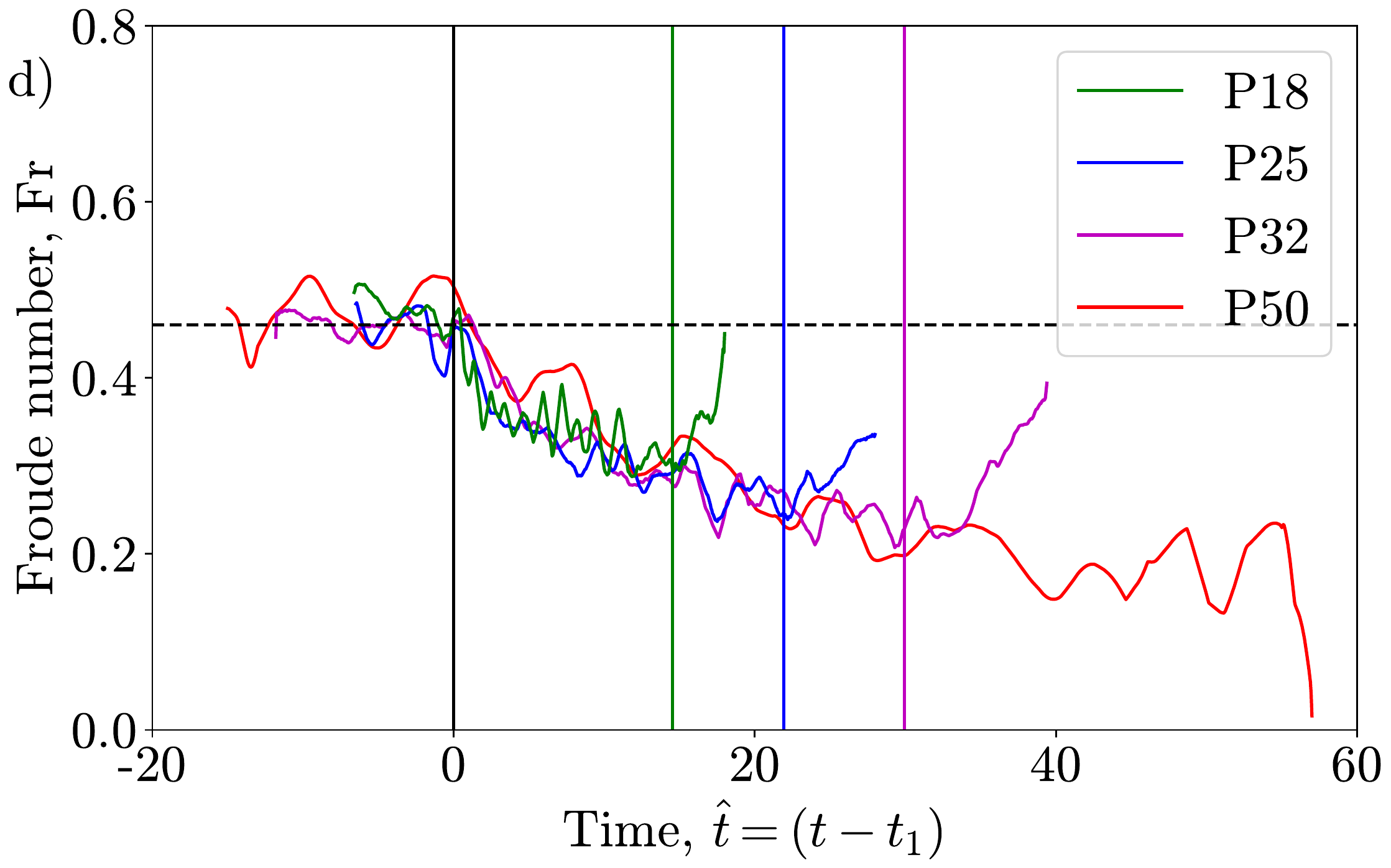}
\end{tabular}
	\caption{Front locations (a, b) and Froude numbers (c, d) for sparse (a, c) and plunging (b, d) experiments with time. Froude numbers measured by \cite{cenedese2018} are marked on panel c by circles with the total measurement window shown by the horizontal black lines. The dashed line on panels c and d shows the average Froude number before encountering the roughness array. The vertical lines on panels c and d show the time when each current exited the roughness array.}
	\label{fig:exp:posvelo}
\end{figure}

It has been shown that similar roughness arrays cause the Froude number of a gravity current to decrease \citep{cenedese2018}.
However, due to measurement limitations, previous studies have typically assumed that the currents were in a quasi-steady state.
Fig.~\ref{fig:exp:posvelo} presents the front position (top row) and the Froude number (bottom row) as a function of time for all sparse (left) and plunging (right) experiments.
As noted in \S\ref{sec:ND}, within our non-dimensional framework the front velocity and the Froude number are equivalent.
The larger measurement window in our experiments, as compared to that used in \cite{cenedese2018}, allows us to calculate the Froude number without the assumption that the current is in a quasi-steady state and hence, determine how the Froude number changes with time.
The vertical lines on Fig.~\ref{fig:exp:posvelo} show the times when the current exited the roughness array.
Due to the non-dimensionalisation, these times differ for each experiment despite the dimensional length of the roughness array being constant across all experiments.


Periodic fluctuations are present in the Froude numbers shown in panels~(c) and~(d) due to the seiche waves that are inevitably produced by the removal of the lock gate \citep{kollner2020gravity}.
Prior to the roughness ($\hat{t}<0$), all currents behave as canonical smooth-bed currents with a Froude number of approximately 0.46 as shown by the dashed line on Figs.~\ref{fig:exp:posvelo}(c) and (d) \citep{huppert1980slumping,Britteranote,sherLockExchange}.
Once the currents enter the roughness ($\hat{t}=0$), they immediately begin to decelerate.
While in the roughness array, the currents never appear to reach a steady state and instead continually decelerate.
Experiments with roughness in the sparse configuration and the plunging configuration show the same general behaviour although the rate of deceleration appears to be larger for the plunging configuration.

The deceleration observed in Fig.~\ref{fig:exp:posvelo} is independent of the roughness height.
However, interpretation of this result is complicated by the {dimensional} rather than the non-dimensional cylinder spacing being constant across experiments.
During one non-dimensional time or {distance}, gravity currents with a smaller non-dimensional cylinder height will encounter more rows of roughness than those with a larger non-dimensional cylinder height.
Equivalently, all experiments contain the same number of rows of roughness but the final Froude number at the end of the roughness array is significantly lower for experiments with high non-dimensional roughness heights than for those with small roughness heights.
Nonetheless, it is noteworthy that, within our non-dimensional framework, the rate of deceleration within the roughness array is the same for all experiments.


Measurements of the Froude number presented in \cite{cenedese2018} were presented based on a 500\,mm long measurement window that was centered 1000\,mm downstream of the start of the roughness.
The reported Froude numbers are shown on Fig.~\ref{fig:exp:posvelo}(c) at the non-dimensional time when the currents in this study had travelled 1000\,mm and the black lines show the range of times where the current front in this study was within the measurement window in \cite{cenedese2018} (i.e. 750--1250\,mm from the start of the roughness).
The instantaneous Froude numbers measured in this study are consistent with those reported in \cite{cenedese2018} and the change in Froude number across the 500\,mm long measurment window is seen to be small.
However, it is clear that the currents in our study were still decelerating despite the assumption of quasi-steady state made in \cite{cenedese2018}.


\begin{figure}
\centering
\includegraphics[width=0.6\textwidth]{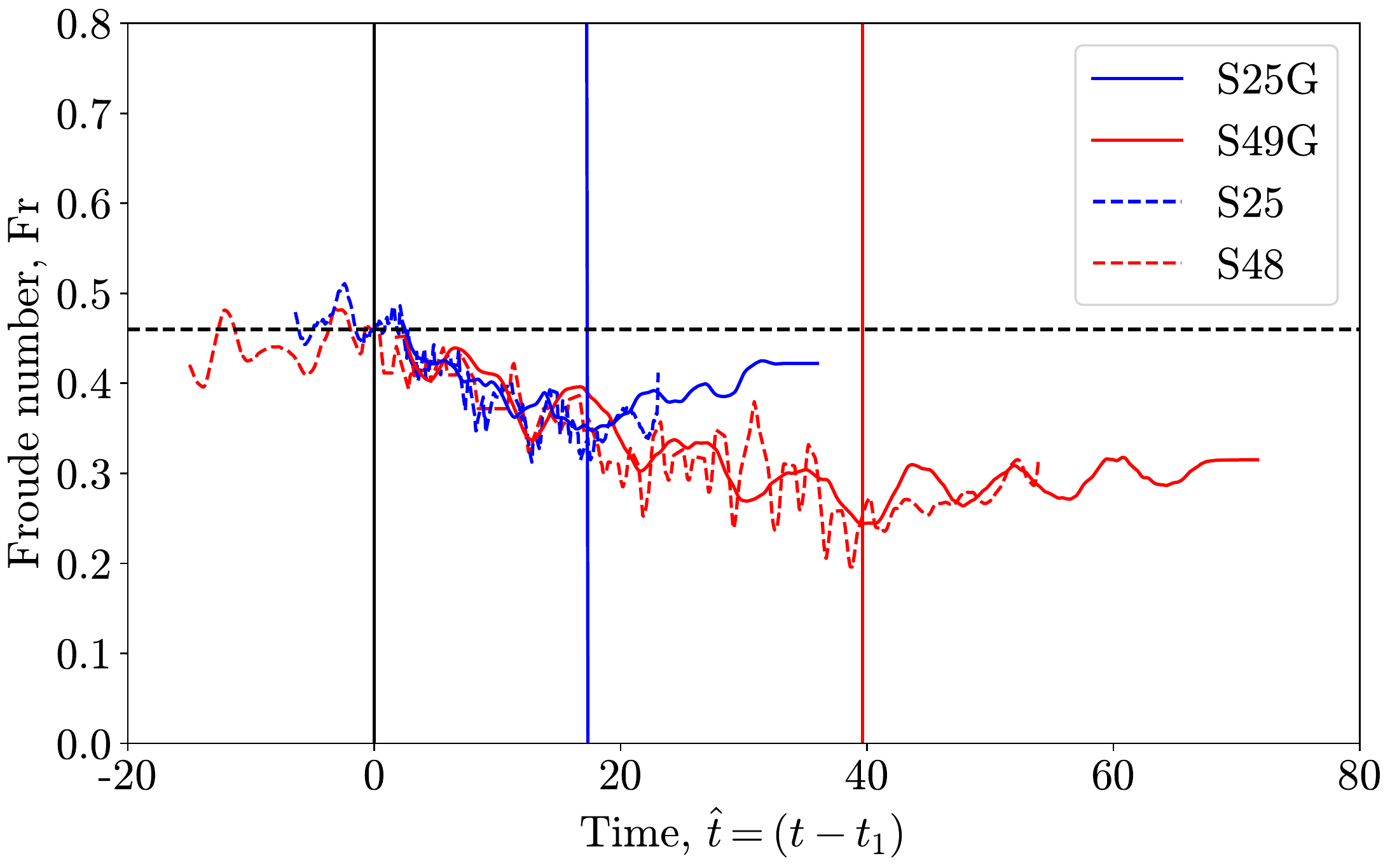}
\caption{
Froude number as a function of time for experiments with the roughness array starting at the gate and experiments with the roughness array starting downstream. The horizontal dashed and vertical lines have the same meaning as on Fig.~\ref{fig:exp:posvelo}.}
	\label{fig:sparsecomp}
\end{figure}

Fig.~\ref{fig:sparsecomp} shows the Froude numbers for the standard experiments and the gate experiments.
For clarity, only two roughness heights are shown.
From Fig.~\ref{fig:sparsecomp} it is clear that the initial development as a smooth bed current has a negligible impact on the downstream behaviour of the current.
This result is also supported by the equal Froude numbers measured in this study and those measured in \cite{cenedese2018} as the roughness array in the latter study started immediately from the gate.

\begin{figure}
    \centering
    \includegraphics[width=0.6\linewidth]{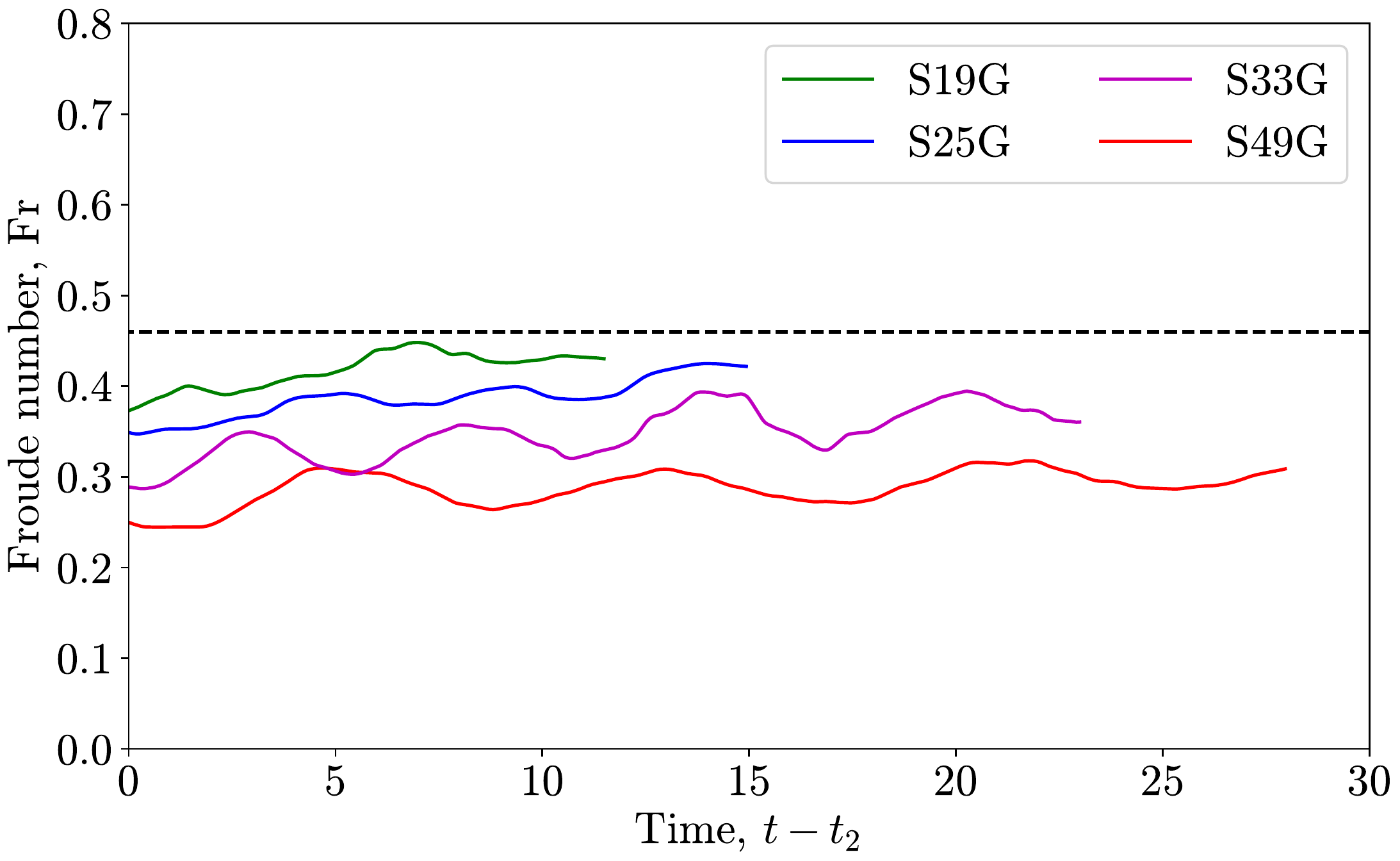}
    \caption{Froude numbers for sparse configuration experiments after emerging from the roughness array. $t=t_2$ refers to the time when each current first exited the roughness.}
    \label{fig:SparseAfter}
\end{figure}

The vertical lines in Figs.~\ref{fig:exp:posvelo} and~\ref{fig:sparsecomp} show the times when currents exited the roughness array.
Beyond this time the currents are seen to accelerate.
The acceleration appears to be equivalent in the standard and gate configurations.
Fig.~\ref{fig:SparseAfter} shows the measured Froude numbers for the gate experiments only after the currents have left the roughness array.
The gate experiments are used because they have a larger measurement window after the roughness array, with Fig.~\ref{fig:sparsecomp} showing that the current behaviour after exiting the roughness is independent on where the roughness started.
For better comparison, {results are plotted as a function of $ t-t_2 $, where $ t_2 $ is the time when the front of the current exited the roughness array.}
The currents are observed to accelerate with an approximately equal acceleration.
However, after some time, the Froude numbers appear to reach a quasi-steady value that is significantly less than the smooth bed Froude number of approximately 0.46.
The quasi-steady Froude number is lower for experiments with large roughness heights than those with small roughness heights.
Considering the balance within the head of a gravity current proposed by \cite{sherLockExchange} where dilution in the head is balanced by replenishment from the tail, it is likely that, after exiting the roughness array, the drag acting on the current is reduced, increasing the replenishment of dense fluid from the tail, causing the head to become denser and the current to accelerate.
Such a process would be independent of the roughness array leading to the uniform acceleration across the experiments.



\subsection{Current height and density}
\label{Sec:Height}

Section~\ref{sec:Qualitative} showed that as the current travels through the roughness array, it transitions from the head and tail structure of a smooth bed current to a wedge structure.
We will now consider the shape of the current in more detail by examining the buoyant height, $h_B$, and the current envelope, $h_c$.
In addition, the ratio between these two terms will be used to provide a measure of the average density in the current.
If the density profile within the current was top-hat (i.e. a uniform density within the current and a uniform density outside the current), the ratio $h_B/h_c$ would give the top-hat current density.
However, as the true density profile within the current becomes stably stratified, the ratio will give an underestimation of the average current density.

{To illustrate the relationship between the density structure of the current and $ h_B/h_c $, \added{Fig.~\ref{fig:DensityProfiles} shows four vertical density profiles from the S25 experiment and three vertical density profiles from the P25 experiment.}\deleted{four vertical density profiles are shown on Fig.~\ref{fig:DensityProfiles}.}
\added{The density profiles relate to panels (c) and (d) of Figs.~\ref{fig:S5H20} and~\ref{fig:P5H20}. 
Only three profiles are shown from the P25 experiment since the location $ \hat{t}=12 $, $ x_F -x = 4$ was upstream of the roughness array.}
\deleted{The density profiles are taken from the S25 experiment shown on Fig.~\ref{fig:S5H20} and the two times correspond to panels c and d of that figure.}
Solid lines show the vertical density profile located 1 and 4 non-dimensional distances behind the front of the current.
Dashed lines show the top-hat profile that has a density of $ h_B/h_c $ from $ y=0 $ to $ y=h_c $ and zero elsewhere.
Fig.~\ref{fig:DensityProfiles} shows that, particularly at a non-dimensional distance of 4 behind the current front, the top-hat density reasonably approximates the true current density structure and that $ h_B/h_c $ is an appropriate measure of the current density.
Trends in $ h_B/h_c $ as the current develops will be discussed later in this section.
	
	\begin{figure}
		\centering
			\begin{tabular}{cc}
				\includegraphics[width=0.48\textwidth]{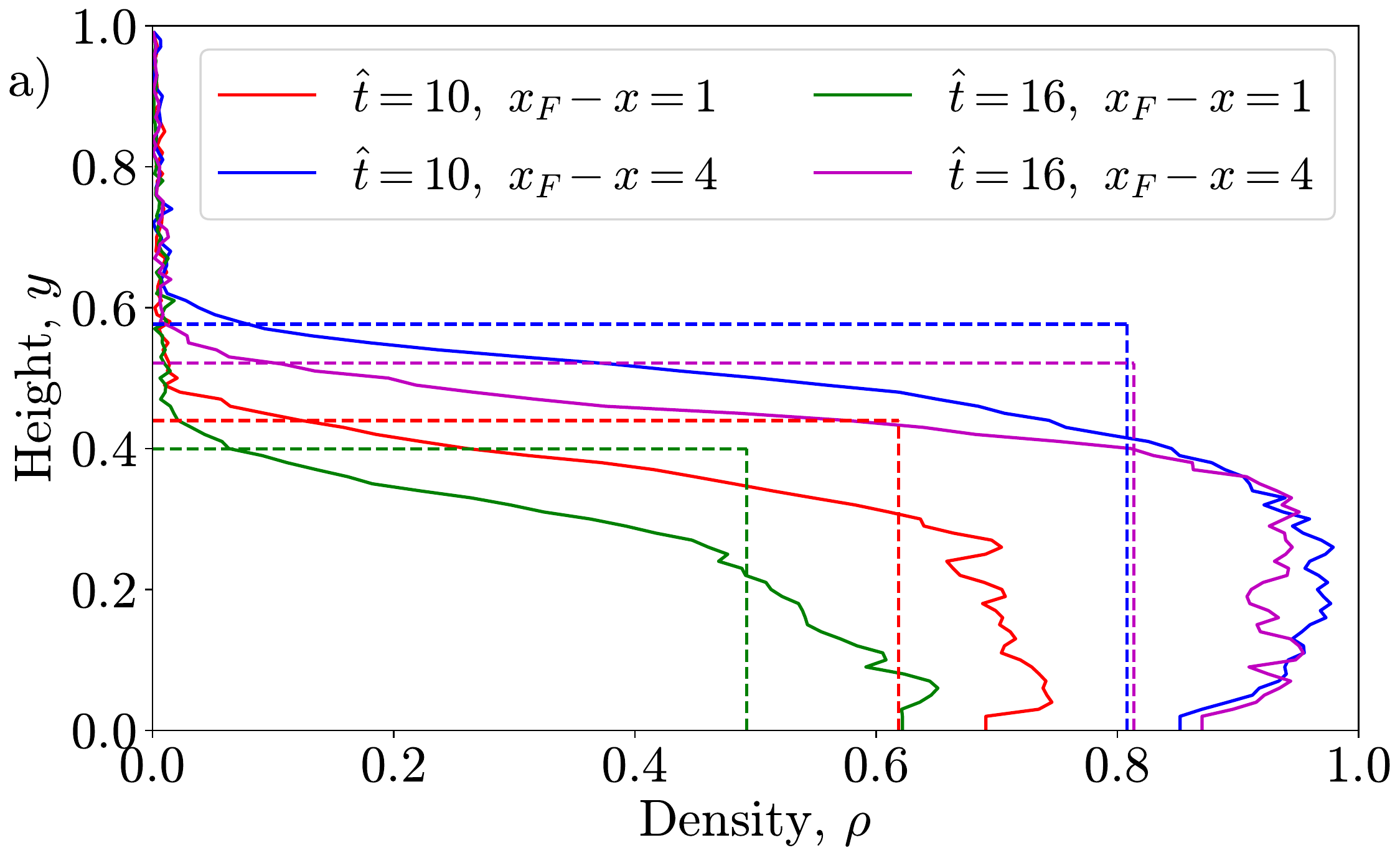}&
				\includegraphics[width=0.48\textwidth]{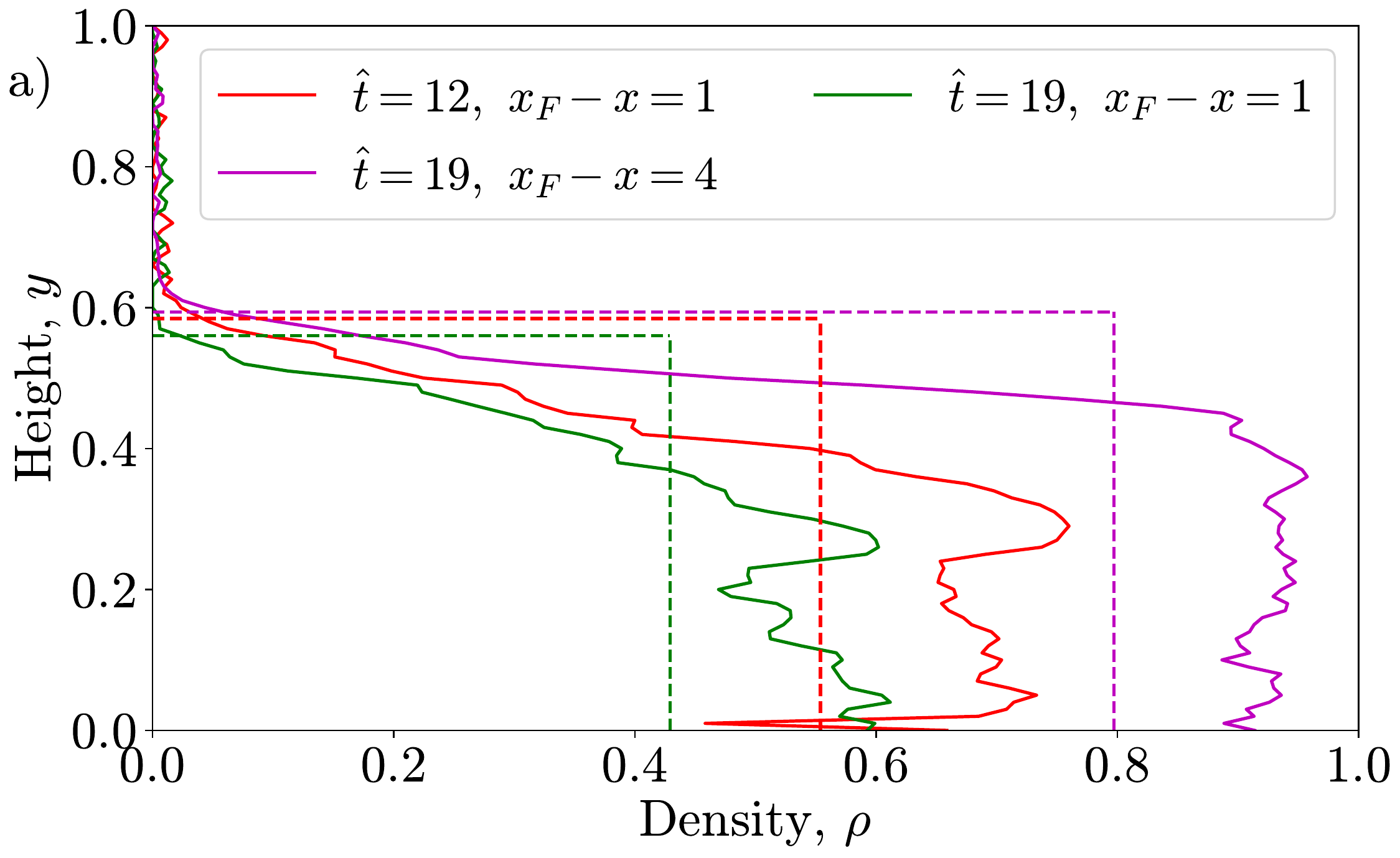}
			\end{tabular}
		\caption{\added{Vertical density profiles (solid lines) and $ h_B/h_c $ at two locations and times for (a) the S25 experiment shown on Fig.~\ref{fig:S5H20}, and (b) the P25 experiment shown on Fig.~\ref{fig:P5H20}. 
			Solid lines show the measured vertical density profiles and dashed lines show the top-hat model of the current with a density of $ h_B/h_c $ to a height of $ h_c $.  Plunging data is not shown for $ \hat{t}=12 $ and $ x_F-x=4 $ as that location was upstream of the current encountering the roughness array.}}
		\label{fig:DensityProfiles}
	\end{figure}

Fig.~\ref{fig:exp:ShinsWithx} shows the buoyant height as a function of $x$ for the sparse and plunging configuration experiments.
The data are plotted at $\hat{t}=11$ for the sparse configuration and at $\hat{t}=14$ for the plunging configuration.
These times correspond to the currents with the smallest roughness heights (S18 and P18) reaching the final row of roughness elements.
Thus, all currents are fully contained within the roughness array.
{Results are plotted based on the position behind the current front, $ x-x_F $} and the vertical line shows the upstream end of the roughness array.
Because the currents all travel at the same speed (Fig.~\ref{fig:exp:posvelo}), the upstream end of the roughness array is at the same location for all experiments within the non-dimensional framework.


\begin{figure}
\centering
\begin{tabular}{cc}
\includegraphics[width=0.48\textwidth]{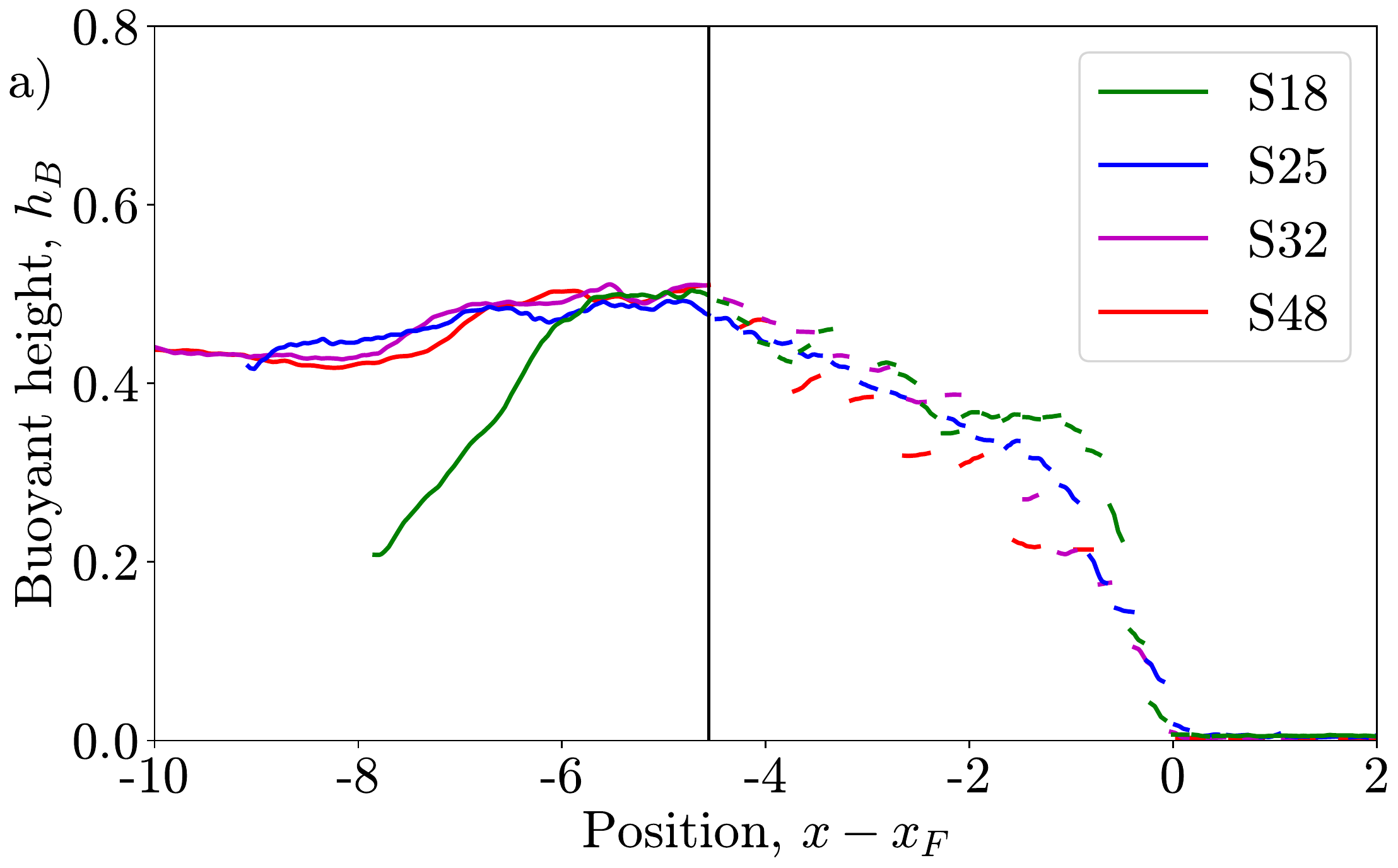}&
\includegraphics[width=0.48\textwidth]{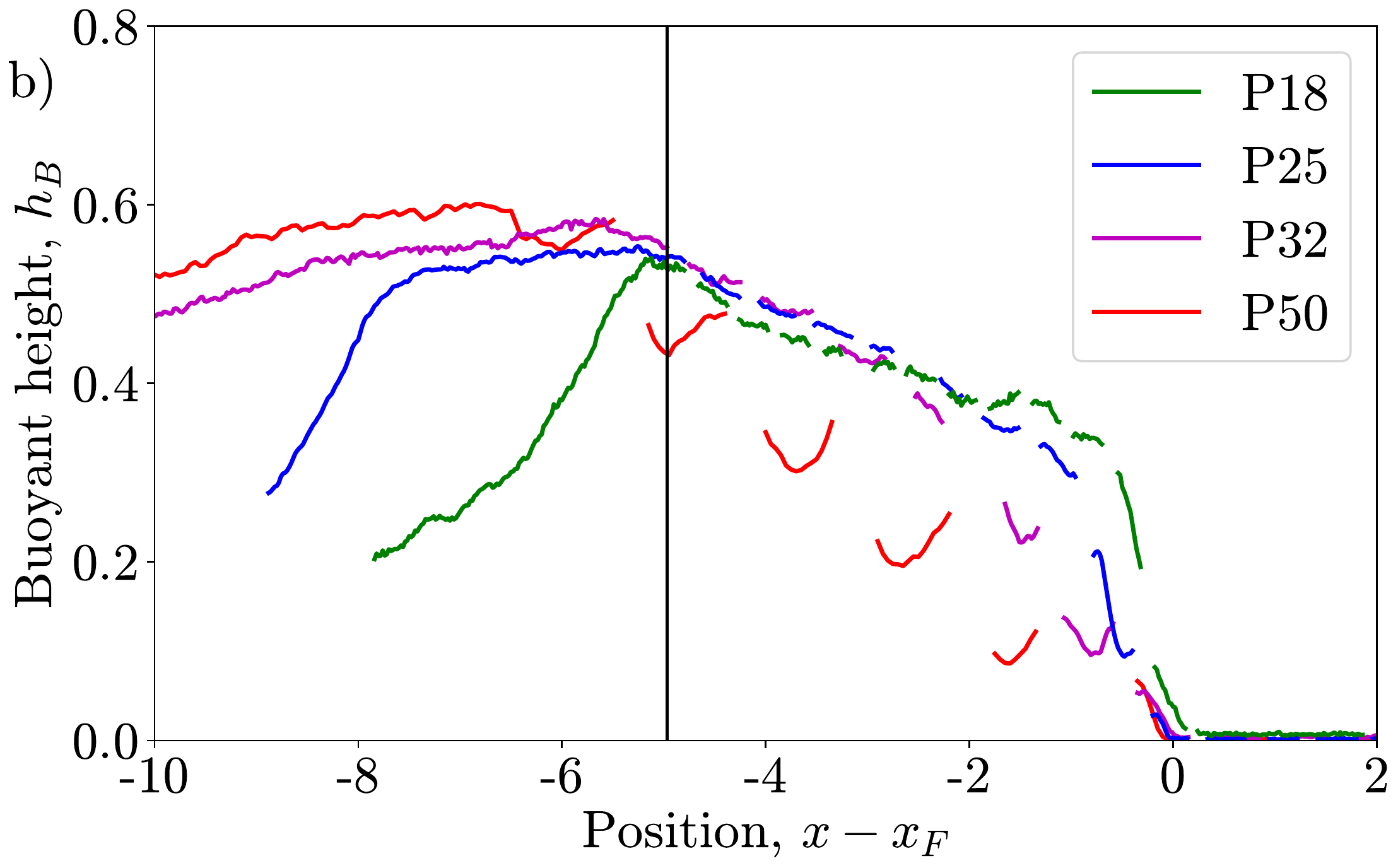}

\end{tabular}
	\caption{Buoyant height, $h_B$, for sparse (a) and plunging (b) experiments as a function of $x$ at the time when the current in experiment S18 and P18 reached the final row of roughness ($\hat{t}=11$ and $\hat{t}=14$, respectively). $x_F$ refers to the current front at the time when measurements were made.}
	\label{fig:exp:ShinsWithx}
\end{figure}

The profile of $h_B$ near the front of the current was slightly dependent on the roughness height.
However, only the two experiments with the smallest roughness height (S18 and P18) had a clear head behind the front.
Within the tail, all experiments except for P50 formed a wedge structure.
\added{Here we define the tail of the current, and the location of the wedge, as $ (x-x_F) < -1 $ based on the the data shown in Fig.~\ref{fig:exp:ShinsWithx} and the observation that the head of a smooth bed current extends approximately one non-dimensional distance behind the front.}
For each of these experiments the buoyant height gradient within the tail was equal and the buoyant height tended to be maximum at the upstream end of the roughness {array}.
{In comparison, a gravity current propagating along a smooth bed with a typical head and tail structure would be expected to have a constant buoyant height within the tail.
The existence of a buoyant height gradient within the tail is therefore an appropriate diagnositic to determine if a gravity current has transitioned to a wedge structure.}

Figs.~\ref{fig:P5H20Plunging} and~\ref{fig:PlungingBores} showed two different types of behaviour within the plunging roughness configuration.
The first, shown on Fig.~\ref{fig:P5H20Plunging} shows the current being deflected over the roughness before plunging over the downstream side.
Fig.~\ref{fig:PlungingBores} shows a similar vertical deflection but, due to the increased roughness height, the current is no longer able to plunge over the rows of roughness and instead flows through the gaps between roughness elements.
These two different behaviours are also apparent on Fig~\ref{fig:exp:ShinsWithx}(b). 
Buoyant height profiles for experiments P18 and P25 are similar to the sparse configuration experiments with a poorly defined head followed by a constant gradient in the tail.
The buoyant height gradient in the tail is equal for the two experiments and similar in magnitude to the sparse configuration experiments.
In contrast, buoyant height profiles for experiments P32 and P50 are elevated immediately upstream of each row of roughness elements.
This effect is most prominent for experiment P50 due to the larger roughness height.
For experiment P32, the effect of each row of roughness ceases to be visible once the buoyant height exceeds the height of the roughness, at which point the current develops the same wedge shape as the sparse configuration experiments.
Thus, whether the current responds to individual rows of roughness or to the entire roughness array appears to depend on whether the buoyant height is less than or greater than the roughness height.


Fig.~\ref{fig:exp:shinTime} shows the development of the current structure with time as it moves through the roughness array.
The top row shows the buoyant height, $h_B$, the second row shows the current envelope, $h_c$, and the bottom row shows the ratio of these two quantities which, as was described at the start of this section, provides an estimate of the average current density.
All of these quantities are measured one non-dimensional distance behind the front of the current which we refer to hereafter as the {standard head location}.
This location corresponds to the location of the head in a smooth bed current, as defined by the maximum value of $h_B$.
{$t_1'$ represents the time when the standard head location reaches the roughness array, rather than the front of the current.
The vertical lines on Fig.~\ref{fig:exp:shinTime} show the times when the standard head location exits the roughness array.}


\begin{figure}
\centering
\begin{tabular}{cc}
\includegraphics[width=0.48\textwidth]{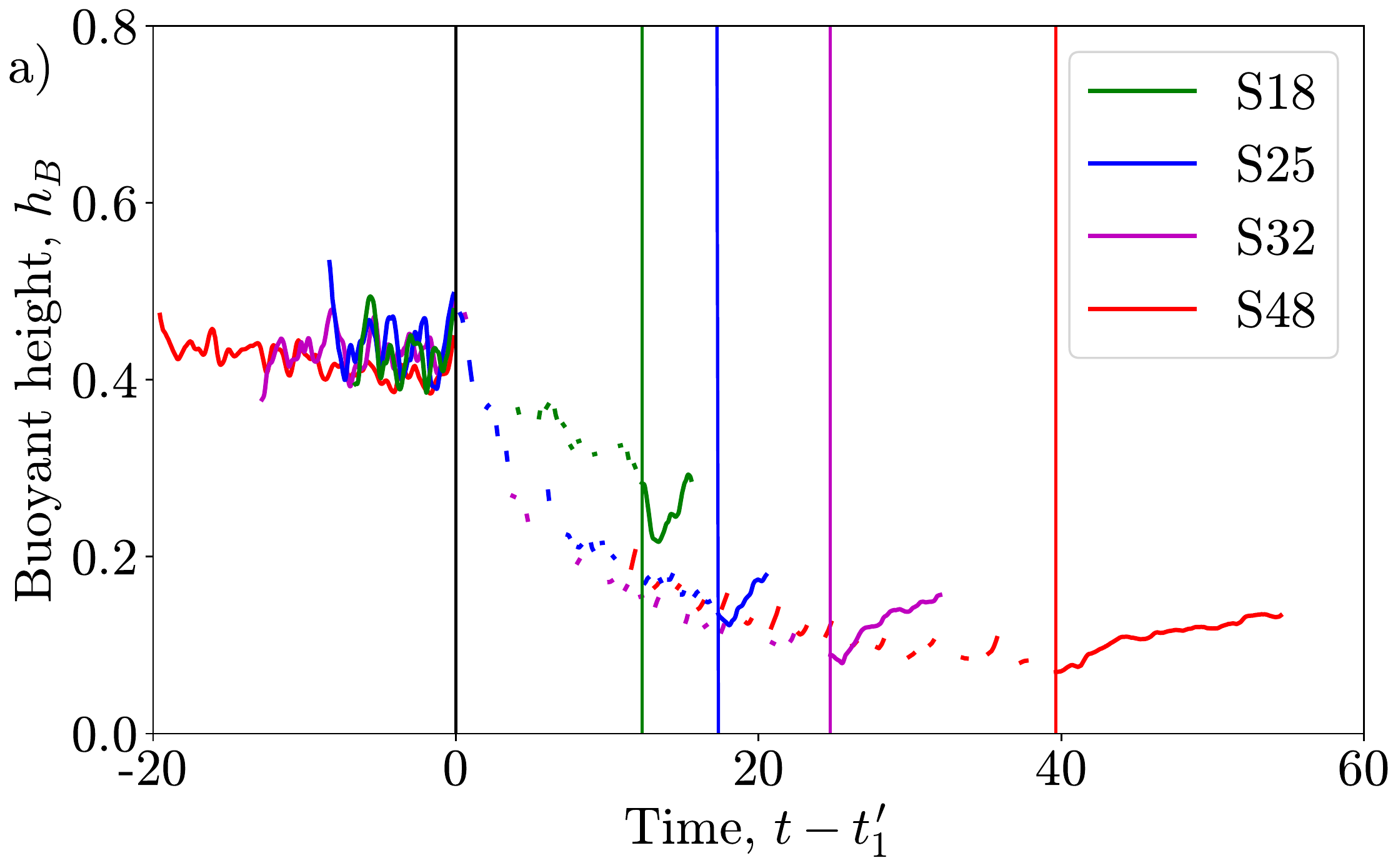}&
\includegraphics[width=0.48\textwidth]{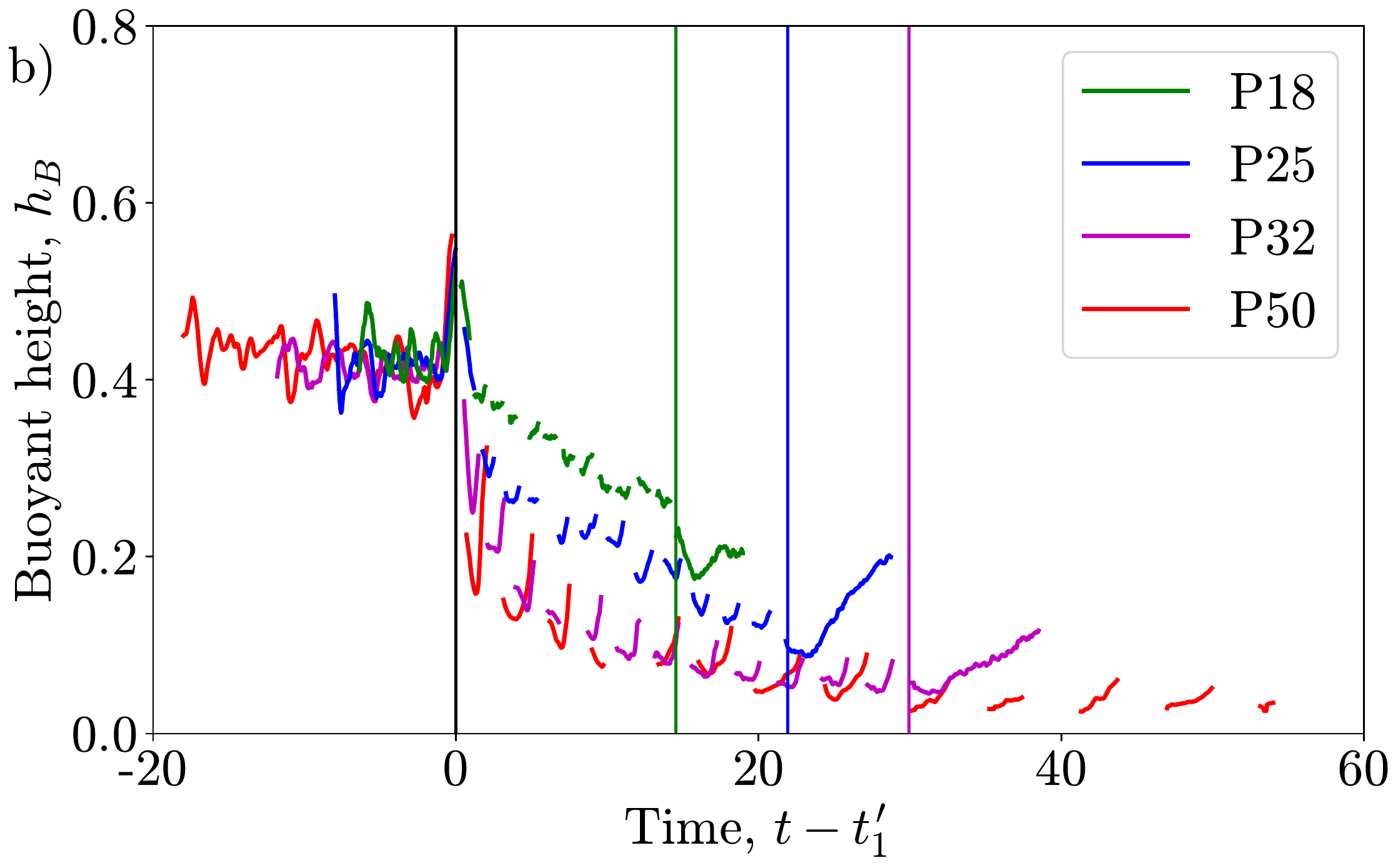}\\
\includegraphics[width=0.48\textwidth]{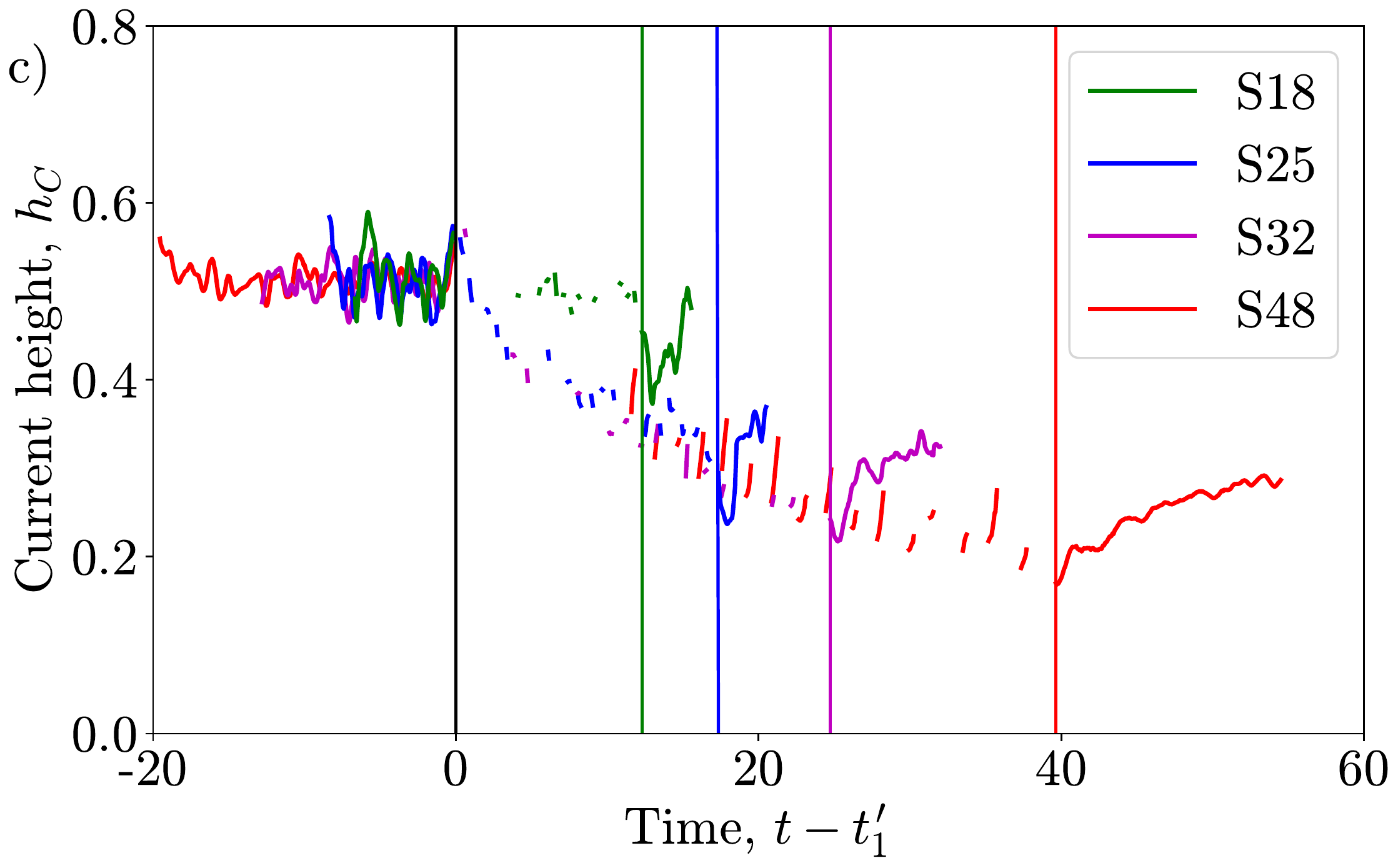}&
\includegraphics[width=0.48\textwidth]{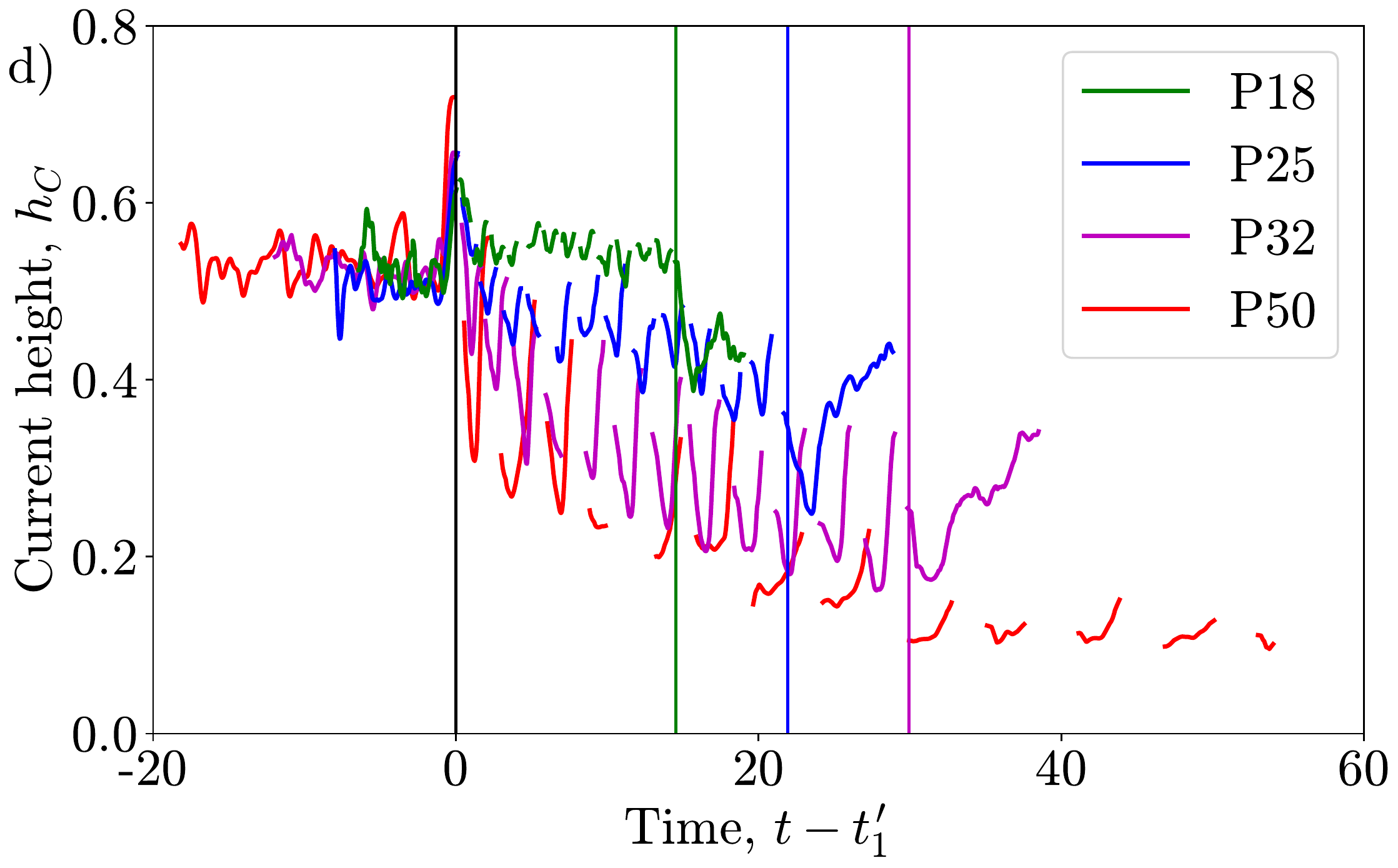}\\
\includegraphics[width=0.48\textwidth]{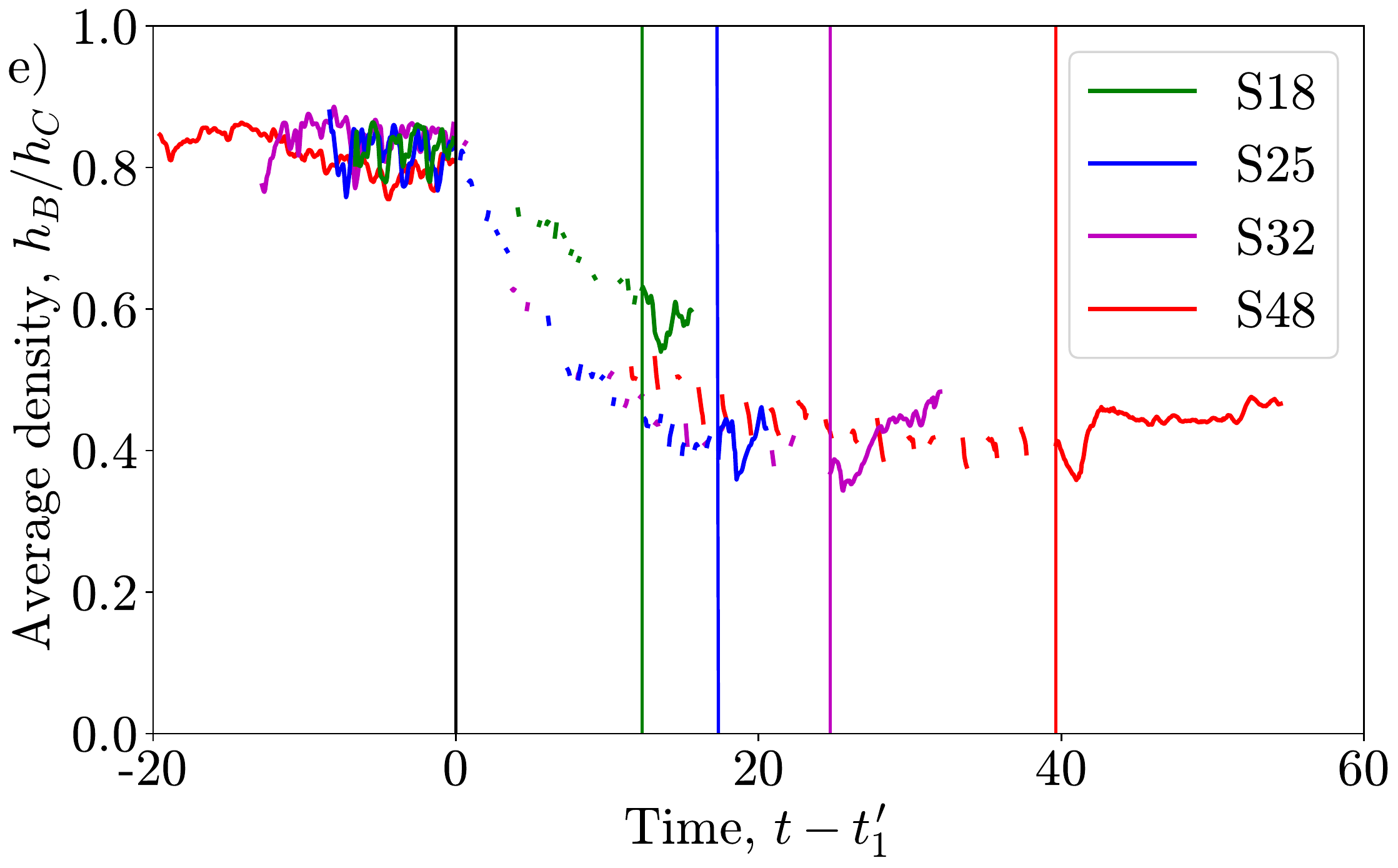}&
\includegraphics[width=0.48\textwidth]{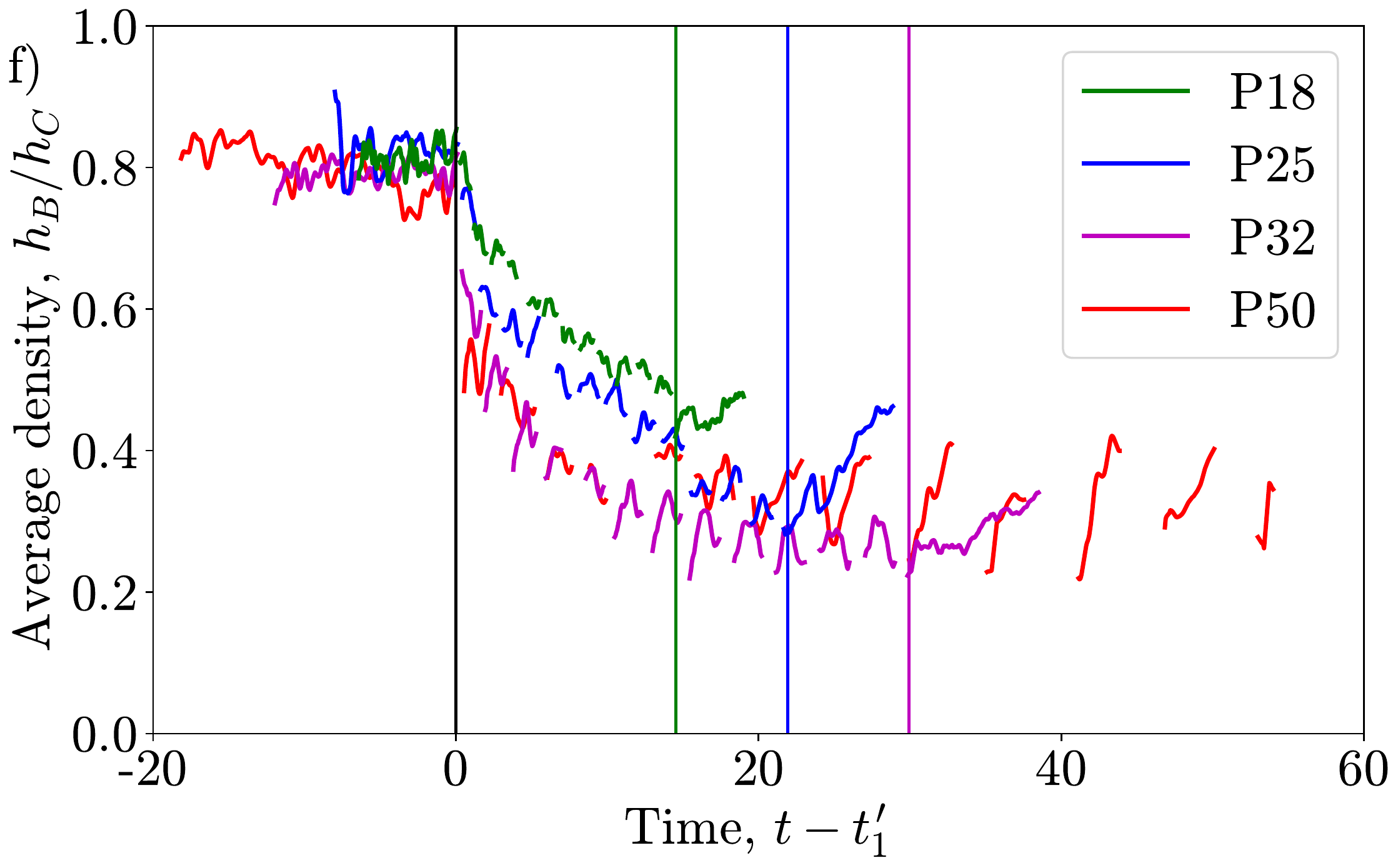}

\end{tabular}
	\caption{Buoyant height (a, b), current envelope (c, d) and an estimate of the average density (e, f) with time. The left column shows sparse configuration experiments and the right column shows plunging configuration experiments. Data are measured a distance of 1 behind the front of the current. As before, the vertical lines show the time when each current exited the roughness array.}
	\label{fig:exp:shinTime}
\end{figure}

Before encountering the roughness array ($t_1'<0$), the currents all behaved as smooth bed gravity currents and had equal properties.
After entering the roughness, the buoyant height, the current envelope, and the average density all decreased with time.
Properties for the sparse configuration experiments decreased at the same rate for all experiments except for S18 which retained properties more similar to a smooth bed current.
Experiment S18 behaving more like a smooth bed current at the standard head location is consistent with the spatial profiles of buoyant height, shown on Fig.~\ref{fig:exp:ShinsWithx}, which included a significant head structure for experiment S18 only.

The plunging configuration experiments show a stronger dependence on the roughness height than the sparse configuration experiments.
The buoyant height and the current envelope both decrease more rapidly for experiments with larger roughness heights.
The reducing buoyant height and current envelope at the standard head location reflects the continual development of the wedge shape.
The buoyant height at the upstream end of the roughness was constant (Fig.~\ref{fig:exp:ShinsWithx}) but the wedge shape was elongated as the current moved through the roughness array, necessitating a decreasing buoyant height gradient and a decreasing buoyant height at the standard head location.

The average current density decreases rapidly with time in both the sparse configuration experiments and the plunging configuration experiments indicating that not only is the wedge shape being stretched, but the current continues to mix with the ambient fluid.
The decrease in density is more rapid for plunging configuration experiments than for sparse configuration experiments, suggesting that the plunging configuration is associated with more rapid mixing.
At late times and particularly for experiments with large roughness heights, the average current density appears to have reached a steady value.
A steady density at the standard head location would suggest that the local balance between mixing of ambient fluid and  replenishment of dense fluid from the tail may be restored \citep{sherLockExchange}.
However, since the current continues to stretch in the flow direction, the buoyant height and current envelope are not expected to reach quasi-steady values at the standard head location.


Considering the plunging configuration experiments shown on Fig.~\ref{fig:exp:shinTime}, a qualitative distinction can be drawn between experiments with large and small roughness heights. 
The currents in experiments P32 and P50 respond to individual rows of roughness with local maxima and minima in the buoyant height and current height which are associated with the roughness element locations.
Focusing on the current envelope, a repeated pattern can be seen where the current is deflected upwards immediately before a row of roughness elements (where there is a gap in the data).
In contrast, the currents in experiments P18 and P25 appear qualitatively similar to the sparse configuration experiments where the current responds to the array of roughness as a field rather than distinct rows of roughness elements.
As a result, the decrease in buoyant height and current envelope tends to be more monotonic with less significant vertical deflections.


After leaving the roughness array, the buoyant height and current envelope were seen to increase for all gravity currents.
The increase in buoyant height, in particular, is consistent with the observation that the gravity currents accelerate after leaving the roughness array as shown on Fig.~\ref{fig:exp:posvelo} and the explanation that there is a net replenishment of dense fluid from the tail of the gravity current.
Although the general trend is for the buoyant height and current envelope to increase after leaving the roughness array, initially a significant reduction is observed (Fig.~\ref{fig:exp:shinTime}).
This reduction can be explained by the current front acceleration due to the absence of drag from the cylinder array happening immediately after the current front leaves the roughness array, while the replenishment of dense fluid from the tail needs some time to restart.

The wedge structure observed in Fig.~\ref{fig:exp:ShinsWithx} suggested that, although there was some variation between experiments in the buoyant height at the standard head location, further upstream all currents had a similar structure independent of the roughness height or configuration.
The behaviour within the tail is further explored in Fig.~\ref{fig:Shin4Behind} where the buoyant height, four non-dimensional distances behind the front, is plotted against time.
{We again define $ t_1' $ so that it still represents the time when the location of interest, in this case four non-dimensional distances behind the front, first encounters the roughness array.}
At the front face of the roughness array, the buoyant height increases as current fluid is blocked by the roughness array, as seen on Fig.~\ref{fig:S5H20}.
With the exception of experiment P50, the buoyant height in the tail then decreases with time after the current enters the roughness array and is the same for all experiments.
As noted above, the wedge angle became shallower over time due to the buoyant height at the start of the roughness array being constant while the current front propagated through the roughness array.
Due to the fixed measurement location relative to the current front, the decreasing wedge angle manifests in Fig.~\ref{fig:Shin4Behind} as a decreasing buoyant height with time.
As has previously been observed, experiment P50 behaves differently due to the greater interaction with each individual row of roughness.


\begin{figure}
\centering
\begin{tabular}{cc}
    \includegraphics[width=0.48\textwidth]{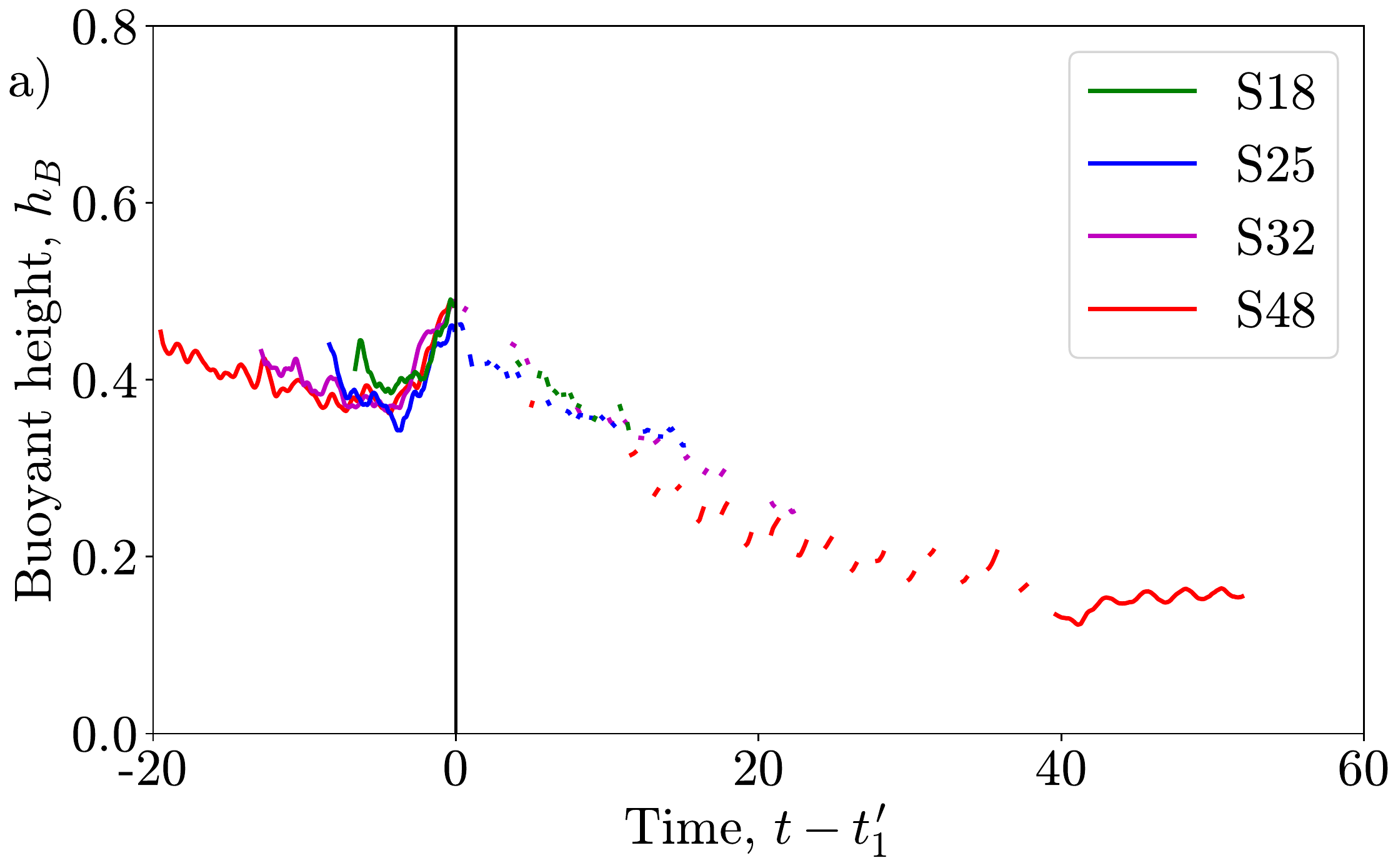}&
    \includegraphics[width=0.48\textwidth]{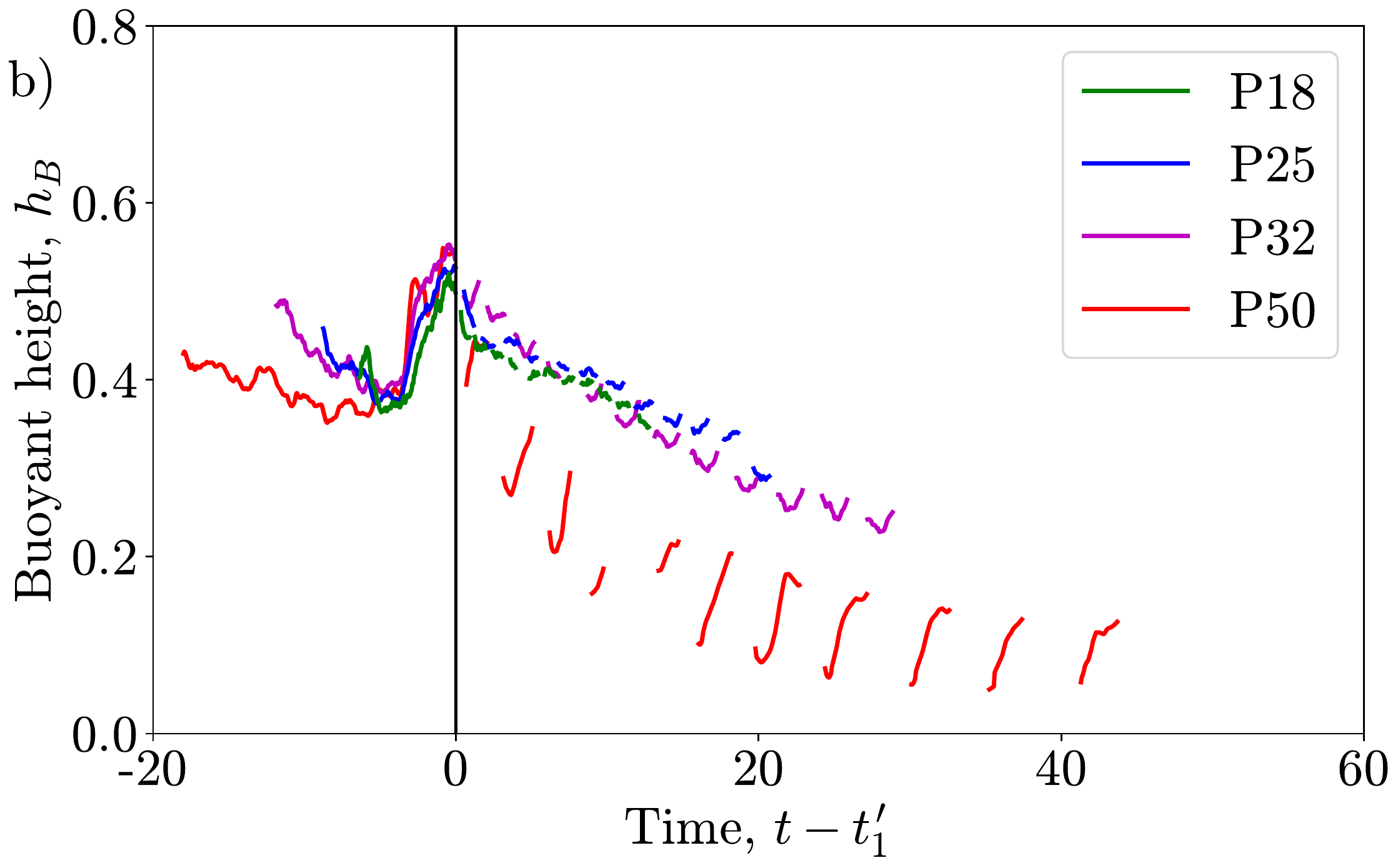}
\end{tabular}
    \caption{Buoyant height a distance of 4 behind front with time for sparse configuration experiments (a) and plunging configuration experiments (b). $t_1'=0$ corresponds to the measurement location first encountering the roughness array.}
    \label{fig:Shin4Behind}
\end{figure}

Fig.~\ref{fig:SparseAfter} showed that the currents accelerated after leaving the roughness array. 
To examine the current structure after leaving the roughness array, Fig.~\ref{fig:ShinAfter} shows the buoyant height and the current envelope at the standard head location as a function of $ t-t_2 $ for the gate experiments.
Both the buoyant height and current envelope increase as the currents exit the roughness array, tending towards steady state properties more typical of smooth bed currents (i.e. $h_b\approx 0.44$ and $h_c\approx 0.5$).
The increase in buoyant height, in particular, is consistent with the re-establishment of a balance between mixing in the head and replenishment of dense fluid from the tail.


\begin{figure}
\centering
\begin{tabular}{cc}
    \includegraphics[width=0.48\textwidth]{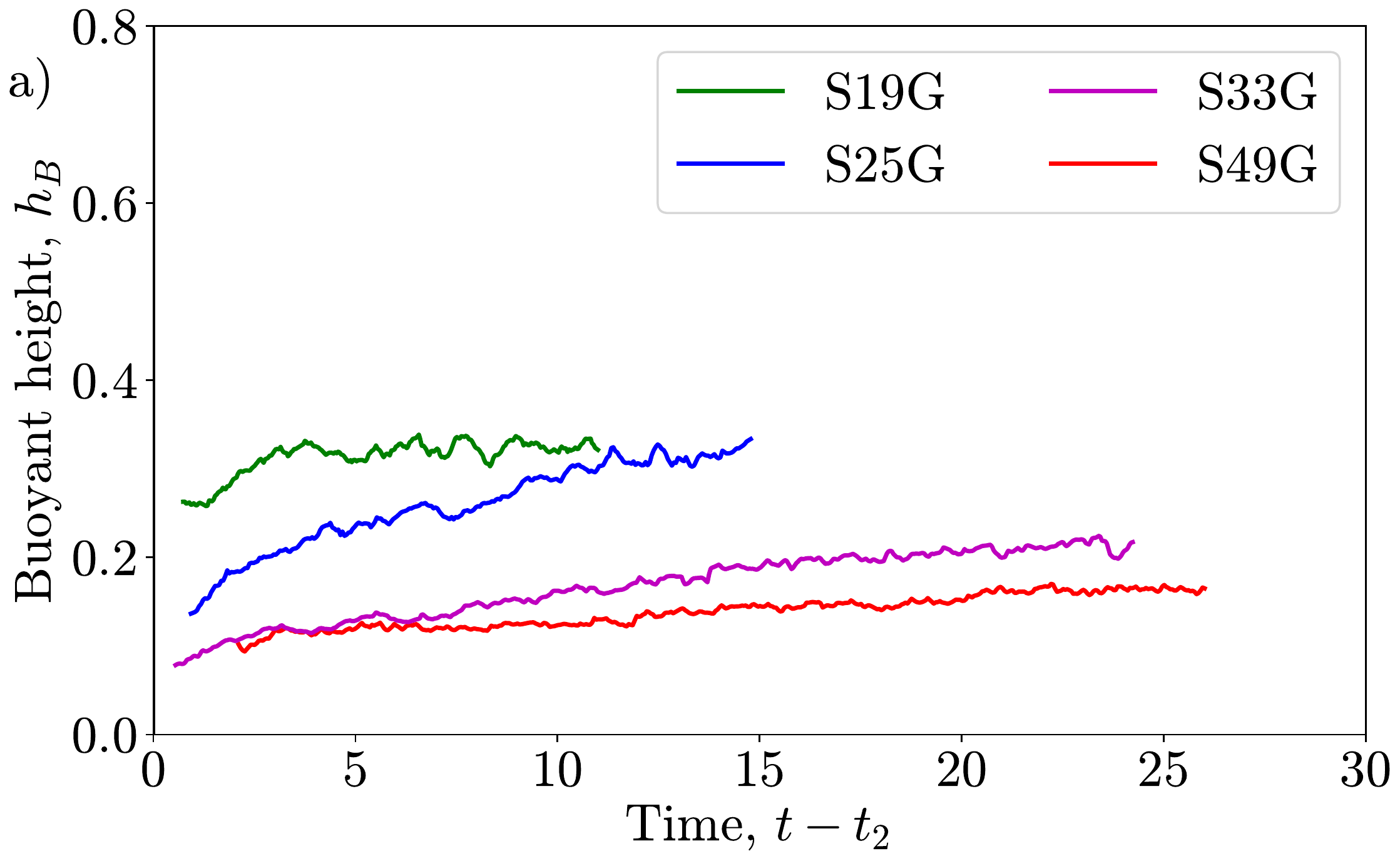}&
    \includegraphics[width=0.48\textwidth]{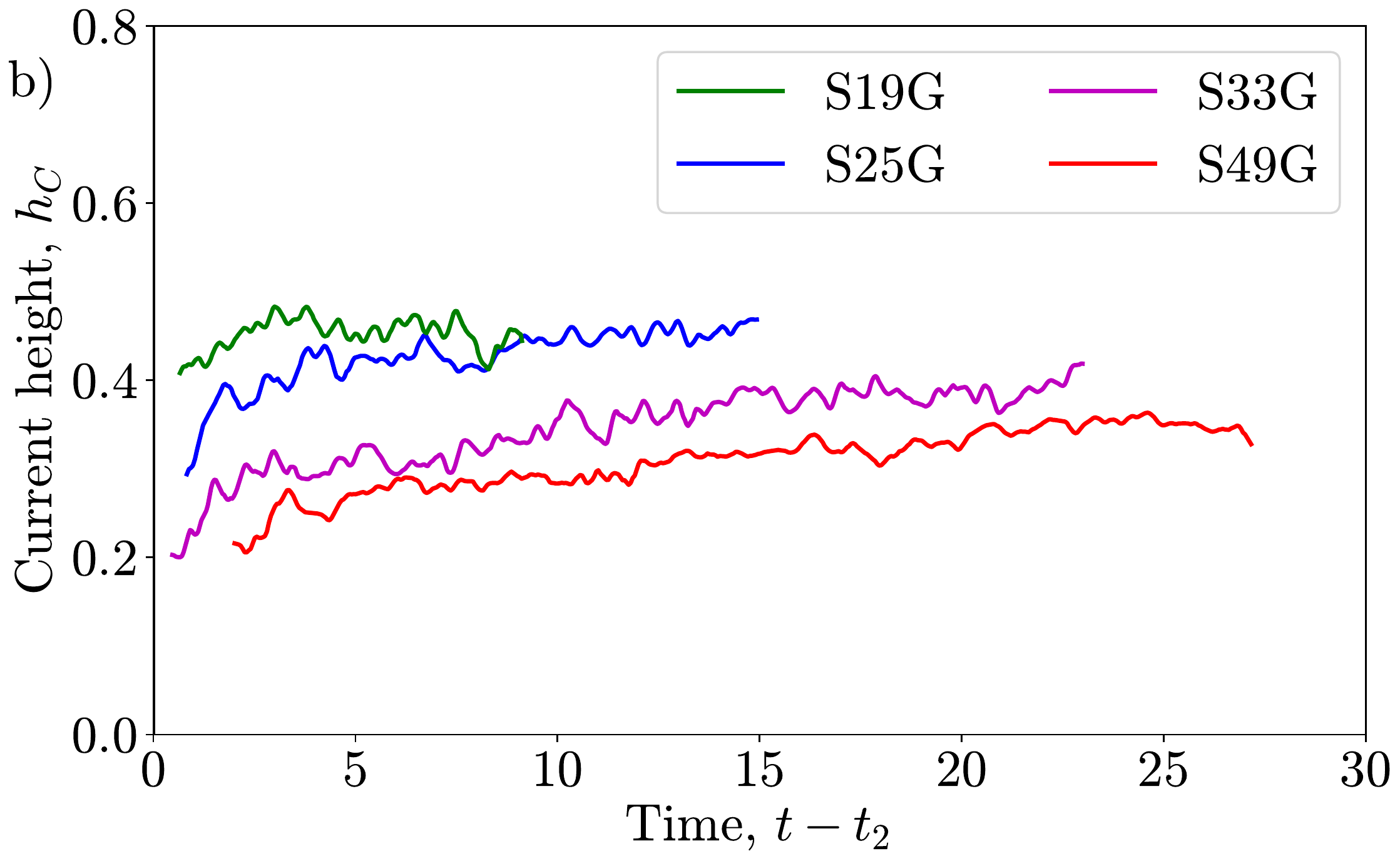}
\end{tabular}
    \caption{$h_B$ (a) and $h_c$ (b) as a function of time for sparse currents at the standard head location after the currents exited the roughness.}
    \label{fig:ShinAfter}
\end{figure}

Experiments with larger roughness heights exit the roughness array with smaller buoyant heights and current envelopes due to spending a longer non-dimensional time within the roughness array (see Fig.~\ref{fig:exp:shinTime}).
Associated with this difference, the experiments with larger roughness heights take a longer time to reach a steady state after leaving the roughness array.
However, it is unclear, particularly regarding the buoyant height, if the steady state properties far downstream will be affected by the roughness height.

\section{Froude number model}
\label{sec:Model}
The experimental results have highlighted that as the currents travel through the roughness array, they tend to transition from the head and tail structure typical of smooth bed currents to a wedge structure.
Furthermore, the wedge structure appears to be independent of the roughness height for most of the experiments (see Figs.~\ref{fig:exp:posvelo} and~\ref{fig:Shin4Behind}).
These observations suggest an alternate dynamic balance to that which is found in smooth bed currents.
Specifically, we propose that the dynamic balance within the roughness array is between a driving pressure gradient in the tail and a drag force exerted by the roughness elements on the tail.
Within this section we formulate a model for the gravity current velocity based on this balance and compare it to the experimentally measured Froude numbers.


Within this description, all quantities will be non-dimensionalised including the cylinder dimensions that were previously described dimensionally.
We will assume that the current consists of an undiluted, linear wedge with a density $\rho_2=1$ which is propagating towards ambient fluid with a density of $\rho_1=0$.
{Although the fluid near the head of the current is observed to dilute relatively quickly as the current propagates through the roughness array, the fluid within the tail remains an approximately constant density (Figs.~\ref{fig:S5H20} and~\ref{fig:DensityProfiles}).
\added{After further development of the current, the assumption of constant density in the tail may become unreasonable.
Nonetheless, for the currents studied here, the current dynamics change due to the lock bore propagating through the tail before there are significant changes of density in the tail.}
Given that the our modelled current is driven by a pressure gradient in the tail, it is reasonable to make the assumption that the current is undiluted.
An undiluted current leads to the buoyant height and current envelope being equivalent, and both will be referred to as $ h $.
For comparison with a real gravity current, $ h $ is better thought of as the buoyant height since the buoyant height  reflects the driving forces within the current.}

Based on Fig.~\ref{fig:exp:ShinsWithx}, we treat the current height at the upstream end of the roughness array as constant and given by $a$.
The height of the current within the roughness array is therefore given by 
\begin{equation}
    h = \left(\frac{x_F-x}{x_F}\right)a,
\end{equation}
where $x_F$ is the non-dimensional front location and $x=0$ corresponds to the upstream end of the roughness array.
Despite $h$ tending to zero at $x_F$, we assume that the roughness elements are always fully submerged within the current.
{The model parameters are shown schematically on Fig.~\ref{fig:ModelSchematic}.}
\added{The model assumptions become increasingly innaccurate near the front of the current where characteristics typical of a current head are observed.
Specifically, the current density is less than the original lock fluid (Fig.~\ref{fig:DensityProfiles}) and the buoyant height gradient is steeper than in the tail of the current (Fig.~\ref{fig:exp:ShinsWithx}).
However, given that we expect the current to be driven by the pressure gradient in the tail, discrepancies at the head will be relatively unimportant to the dynamics of the current.}

\begin{figure}
	\centering
		\includegraphics[width=0.7\textwidth]{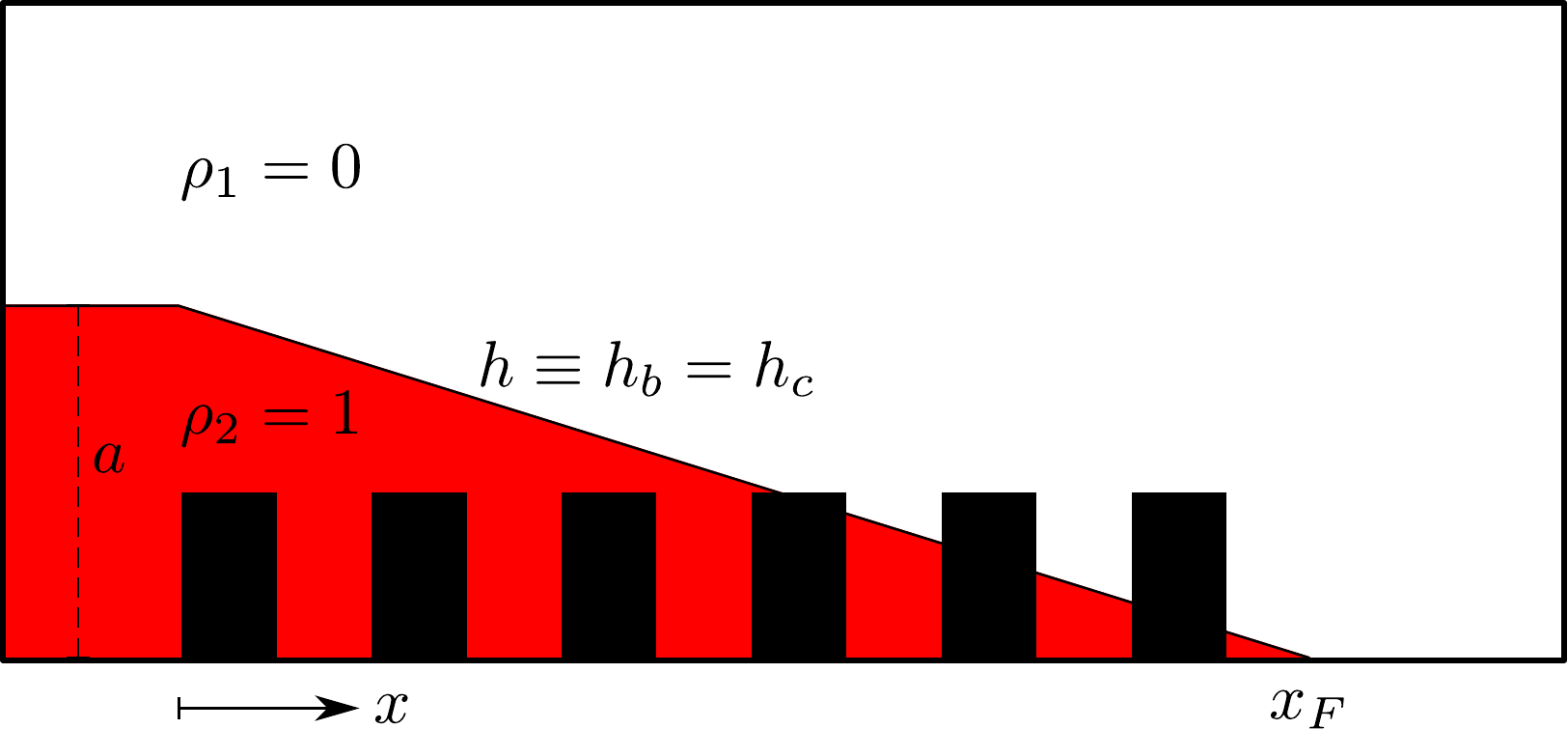}
	\caption{{A schematic of the key model variables. The buoyant height and current envelope are equuivalent due to the fluid within the current being undiluted.}}
	\label{fig:ModelSchematic}
\end{figure}

Assuming that the pressure within the tail is hydrostatic or that the dynamic pressure is slowly varying in the $x$ direction, the non-dimensional horizontal pressure gradient, ${\rm d}P/{\rm d}x$, is given by
\begin{equation}
\frac{{\rm d}P}{{\rm d}x} = \frac{a}{x_F}.
\label{eq:dPdx}
\end{equation}
Individual roughness elements are considered to impose an equal and cumulative drag force on the current which is modelled by a quadratic drag law:
\begin{equation}
    F_D = \frac{1}{2}\,C_d A(\alpha \,{\rm Fr})^2,
    \label{eq:F_d}
\end{equation}
where $C_d$ is the drag coefficient, $A$ is the surface area of a roughness element, Fr is the Froude number of the current (or the front velocity) and $\alpha$ is a {parameter} that relates the front velocity to the internal fluid velocity within the tail.
{The parameter $ \alpha $ is assumed to be constant for a given gravity current, but it is otherwise unknown}.
The vertical surface area (i.e. excluding the top and bottom area) of an individual cylindrical roughness element is
\begin{equation}
    A_o = \pi d_c\eta_c,
\end{equation}
and the total number of roughness elements within the current is
\begin{equation}
    N = \frac{x_FW}{S_x S_z},
\end{equation}
giving the total surface area of the roughness array as
\begin{equation}
    A = A_o N = \frac{\pi d_c \eta_c x_F W}{S_x S_z}.
    \label{eq:A}
\end{equation}
Here $d_c$ is the roughness diameter, $\eta_c$ is the roughness height, $W$ is the width of the channel, $S_x$ is the roughness spacing in the flow direction, and $S_z$ is the roughness spacing across the channel.
All length scales are non-dimensionalised by the total height of the fluid.
From Eqs.~(\ref{eq:F_d}) and~(\ref{eq:A}), the drag force per unit volume acting on the current can be written as
\begin{equation}
    \frac{F_D}{V} = \pi  \left(\frac{d_c \eta_c}{S_x S_z a}\right) C_d \alpha^2\, {\rm Fr}^2.
    \label{eq:FDV}
\end{equation}

Equating the pressure gradient given in Eq.~(\ref{eq:dPdx}) with the drag force per unit volume given in Eq.~(\ref{eq:FDV}), and rearranging for Fr gives
\begin{equation}
    {\rm Fr} = \sqrt{\frac{\mathcal{B} \,\mathcal{C}}{x_F}},
    \label{eq:Fr_x}
\end{equation}
where $\mathcal{B} = a^2/\pi C_d \alpha^2$ combines all numerical constants and $\mathcal{C} = S_x S_z/d_x \eta_c$ describes the geometry of the roughness array.
{$\mathcal{B}$ links the three unknown constants ($ a $, $ C_d $, and $ \alpha $) into one parameter such that they do not need to be independently evaluated from the experimental results.}
Since ${\rm Fr} \equiv {\rm d}x_F/{\rm d}t$, Eq.~(\ref{eq:Fr_x}) can be integrated to provide an expression for the front position as a function of time:
\begin{equation}
    x_F(t) = \left(\frac{3}{2} \left({\mathcal{B \,C}}\right)^{1/2}\,t + x_{F,0}^{3/2}\right)^{2/3}
    \label{eq:xF_t}
\end{equation}
where $x_{F,0}$ is the front position at $t=0$ which is taken here to be the start of the roughness array.

The form of Eq.~(\ref{eq:xF_t}) is equivalent to that found in \cite{Testik16} based on a similarity solution of gravity currents moving through a full depth roughness array.
Specifically, the $x_F\sim t^{2/3}$ relationship shown in Eq.~(\ref{eq:xF_t}) is found when using a quadratic drag law and assuming {that} the current height has a {self-similar} non-dimensional profile as it travels through the roughness array (Class II currents with $\lambda=2$ and $\delta=0$ in their nomenclature).
{Although such currents were predicted to have non-linear height profiles,  the curvature is minor and unlikely to be noticeable in these experiments (see Fig. 6 of \citet{Testik16}).}
\citet{Testik16} note that they are unaware of observations of gravity currents that exhibit this similarity behaviour, highlighting the novelty of these experimental results.

By substituting $x_F$ from Eq.~(\ref{eq:xF_t}) into Eq.~(\ref{eq:Fr_x}), an expression can be found for the Froude number as a function of time, which can then be compared with the experimental data:
\begin{equation}
    {\rm Fr} = {\rm Fr}_0\left(1+\frac{3}{2}\,\frac{{\rm Fr}_0^{3}}{\,{\mathcal{B\,C}}}\,t\right)^{-1/3}.
    \label{eq:Fr_t}
\end{equation}
$\rm{Fr}_0$ is the Froude number at $t=0$ and is taken to be 0.46 based on the Froude number at the start of the roughness array. 
$\rm{Fr}_0$ is expected to be constant for all situations where a smooth bed current encounters a roughness array.
$\mathcal{B}$ depends on the height of the current at the upstream end of the roughness, $a$, the drag coefficient, $C_d$, and the ratio between the internal fluid velocity in the tail and the front velocity, $\alpha$.
The experimental results suggest that $a$ is approximately constant.
However, both $C_d$ and $\alpha$ are likely to depend on the particular roughness configuration (i.e. sparse or plunging) and the shape of the roughness elements.
$\mathcal{C}$ can be calculated based on the geometry of the roughness array and is equal to 4.096 for both the plunging and the sparse configuration experiments presented here.
We note that $\mathcal{C}$ does not depend on the fluid depth, $H$, even when the relevant length scales are non-dimensionalised by $H$.

Values of $\mathcal{B}$ are found for both the sparse configuration and the plunging configuration based on an ordinary least squares fit with the experimental data.
From this analysis it is found that appropriate values are $\mathcal{B}=0.40$ for the sparse configuration and $\mathcal{B}=0.16$ for the plunging configuration.
The experimental measurements suggest that $\mathcal{B}$ is constant for all experiments in a given roughness configuration, suggesting that there is no dependence on the Reynolds number of the flow which would depend on the fluid depth.
Although this is true within our experimental parameter space, it will not necessarily hold at much larger, geophysically relevant Reynolds numbers.
{We also note that although $ \mathcal{B} $ appears to be independent of the Reynolds number, it remains possible that the drag coefficient varies with Reynolds number but that this change is offset by changes in $ \alpha $.}
The smaller value of $\mathcal{B}$ for the plunging configuration experiments implies that the product $C_d\,\alpha^2$ is approximately 2.5 times larger than for the sparse configuration.
We are unable to determine if the difference in $\mathcal{B}$ is predominantly due to the drag coefficient or the velocity ratio.
It is possible that the vertical displacement of the current in the plunging configuration leads to higher energy dissipation and a larger drag coefficient than the sparse configuration regime.
However, the gaps between cylinders in the plunging regime are 12\,mm wide ($0.6\,d_c$) compared to 44\,mm wide ($2.2\,d_c$) in the sparse regime so fluid velocities close to the cylinders are likely to be larger in the plunging regime, increasing $\alpha$.

Fig.~\ref{fig:FroudeModels} shows the measured Froude number as a function of time for the sparse and plunging configurations as well as the model predictions.
The experimental data is the same as that plotted on Fig.~\ref{fig:exp:posvelo}.
Both the sparse configuration model and the plunging configuration model are shown on Fig.~\ref{fig:FroudeModels}(b) to highlight that the different values of $\mathcal{B}$ have a significant impact on the predicted deceleration of the current.
Fig.~\ref{fig:FroudeModels} shows that the model is able to accurately predict the deceleration of the current through the roughness array.
The model has one free parameter, $\mathcal{B}$, which will depend on the roughness configuration and the details of the individual roughness elements.

\begin{figure}
\centering
\begin{tabular}{cc}
    \includegraphics[width=0.48\textwidth]{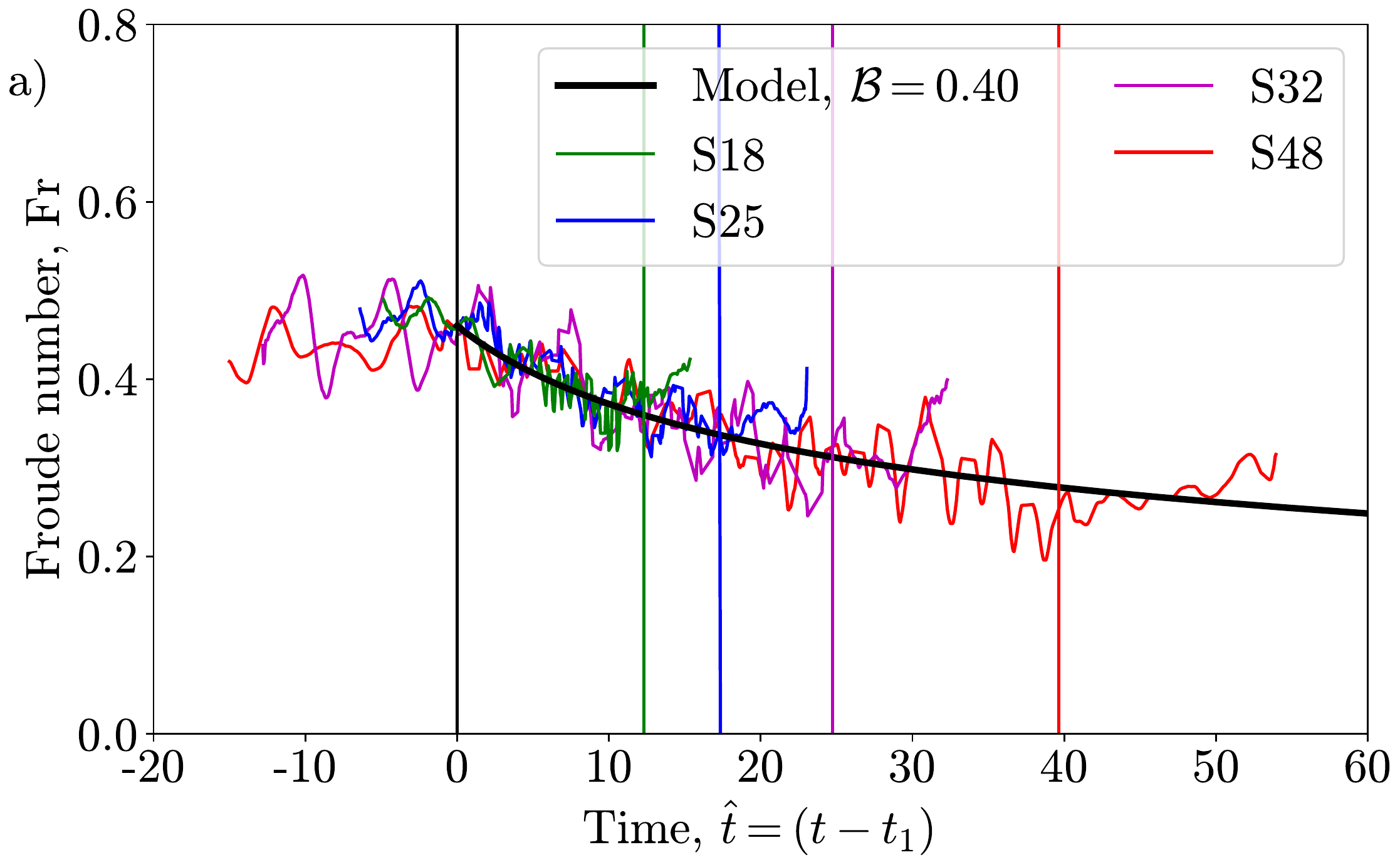}&
    \includegraphics[width=0.48\textwidth]{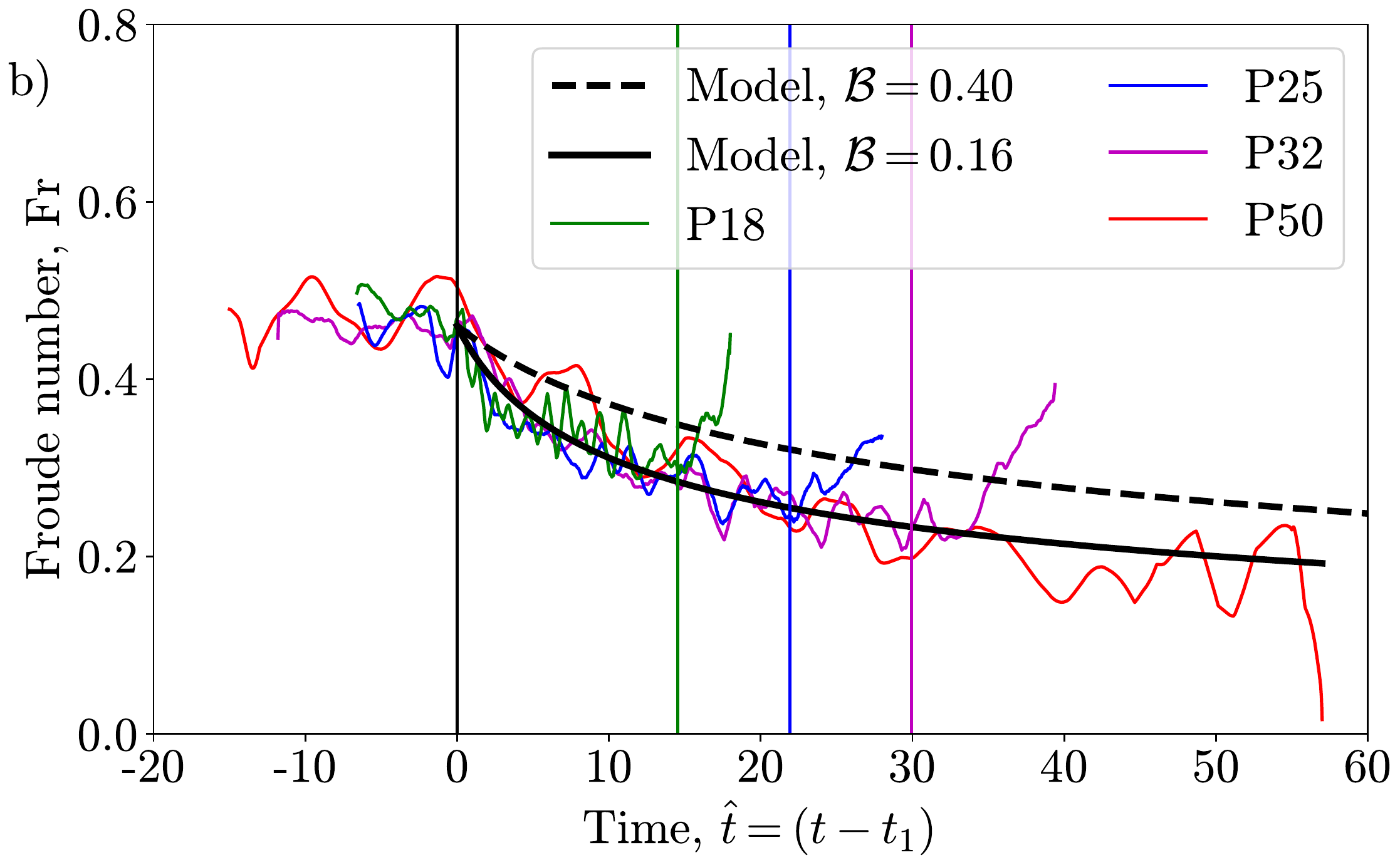}
\end{tabular}
    \caption{Measured and predicted Froude numbers with time for the sparse (a) and plunging (b) roughness configurations. Predicted Froude numbers are based on Eq.~(\ref{eq:Fr_t}) with $\mathcal{B}=0.40$ for the sparse roughness configuration and both $\mathcal{B}=0.40$ and $\mathcal{B}=0.16$ for the plunging roughness configuration.}
    \label{fig:FroudeModels}
\end{figure}
 
Of particular note is that the model does not include any notion of a  frontal region.
It has previously been argued that the frontal region governs the speed of smooth bed gravity currents \citep{nokes2008front}.
Experiments with relatively small roughness heights (e.g.~S18) did display a head region before the wedge structure in the tail.
However, even these experiments demonstrate a close agreement between the model predictions and experimental results.
This supports the assertion that the dynamics of a gravity current travelling through roughness are fundamentally different to that of a smooth bed gravity current and are governed by a balance between the pressure gradient in the tail and the drag force from the roughness elements.


\section{Discussion and conclusion}
\label{sec:conclusion}
The experimental results presented in earlier sections have demonstrated that the dynamics of a gravity current fundamentally change as it enters and travels through an array of roughness elements.
The current changes from the head-and-tail structure that is characteristic of smooth bed currents to a wedge structure.
The model presented in section~\ref{sec:Model} suggests that the change in structure is associated with a change in the dynamic balance from being a balance between buoyancy and inertia in the frontal region to being a balance between the pressure gradient and drag within the tail.
Of particular note, is the consistent behaviour across a range of experimental parameters when viewed within the non-dimensional framework.
For examples, see the Froude number as a function of time (Fig.~\ref{fig:exp:posvelo}), the buoyant height as a function of space (Fig.~\ref{fig:exp:ShinsWithx}), or the buoyant height in the tail as a function of time (Fig.~\ref{fig:Shin4Behind}).
The consistency between experiments highlights that the current behaviour is largely insensitive to the roughness height.

Despite consistent results across the experimental range of roughness heights, the behaviour at more extreme roughness heights remains unclear.
At a sufficiently small roughness height, the influence of the roughness should become insignificant, leading to the current retaining its head and tail structure.
Indications of this are seen for experiment S18 in Fig.~\ref{fig:exp:ShinsWithx} where a clear head is observed.
However, experiment S18 still largely behaves in the same manner as other experiments suggesting that it is still controlled by a pressure-drag balance in the tail.
Determining the maximum roughness height where smooth bed behaviour exists would be a valuable extension to this work.

Similarly, it remains unclear how the current would develop through a longer roughness array.
For all experiments presented here, the current continues to slow and the pressure gradient reduces as the current travels.
It is possible that the observed behaviour will continue until the current becomes sufficiently thin that viscosity becomes important to the force balance.
Alternatively, it is possible that the wedge structure is only a transient response as the current adjusts to the presence of the roughness array.
The roughness array will increase drag within the current, potentially reducing the replenishment of dense fluid into the head that was described by \citet{sherLockExchange}.
This would reduce the buoyancy within the head and slow the current, consistent with the experimental results.
If the wedge structure is a transient response to the roughness array then a head and tail structure could reform further downstream, once the front had slowed sufficiently that mixing within the head was again balanced by replenishment of dense fluid from the tail.

{The} model presented in section~\ref{sec:Model} appears to accurately predict the slowing of the currents with time and is consistent with the earlier model of \citet{Testik16}.
{Our experimental results are also similar in some aspects, such as the current profile, to radial gravity currents from lock-release sources propagating through roughness arrays \citep{Petrolo22}.
These similarities support the experimental results and proposed force balance, and suggest that some results can be generalised to different flow configurations.
Nonetheless, how a gravity current would change in response to different parameter values remains uncertain in a number of cases.}
Across the experiments, the dimensional depth of fluid was varied across two roughness configurations.
Independently testing the dependence of the various length scales ($S_x$, $S_y$, $d_c$, and $\eta_c$ in particular) would provide further confidence in the form of the model and its predictions.
{However, earlier work has shown that the ratio between fluid depth and roughness height is the most important variable when considering the Froude number and structure of gravity currents interacting with roughness arrays \citep{cenedese2018}.
As such, the model is expected to be robust to small changes in other length scales.}
{Separately}, the constant $\mathcal{B}$ is worthy of further investigation to determine how it depends on the roughness configuration. 
$\mathcal{B}$ is, in effect, a drag coefficient based on the front velocity of the gravity current and the experiments showed that it is significantly different between the sparse and plunging roughness configurations investigated here.
Different roughness configurations are likely to have different drag coefficients which would need to be determined from further experimentation.

Finally, previous authors have identified several additional flow regimes when gravity currents encounter roughness arrays \citep{cenedese2018,zhou2017propagation}.
The experiments investigated here have all been within the `through-flowing' or `plunging' regimes and it is likely that `over-flowing' or `skimming' flow regimes would have significantly different dynamics.
Thus, the results should only be applied to situations where the roughness array is relatively sparse.

\begin{acknowledgments}
The authors thank technicians at the University of Canterbury Fluid Dynamics Laboratory for assisting with the experimental programme.
A.\,M. wishes to acknowledge the financial support provided to him through the University of Canterbury doctoral scholarship.
We also thank the anonymous reviewers for their effort and insightful comments in reviewing this paper.
\end{acknowledgments}

\bibliography{ThroughFlowingGravityCurrents}

\end{document}